\documentclass[11pt]{article}
\usepackage{amsfonts, latexsym, amssymb,amsmath}
\usepackage[dvips]{graphics,color}
\newtheorem{theorem}{Theorem}[section]
\newtheorem{definition}{Definition}[section]
\newtheorem{remark}{Remark}[section]
\newtheorem{corollary}{Corollary}[section]
\newtheorem{conjecture}{Conjecture}[section]

\newtheorem{proposition}{Proposition}[section]
\makeatletter 
\@addtoreset{equation}{section}

\makeatother
\begin{document}
\title{Dessins d'Enfants, Their Deformations and Algebraic the 
Sixth Painlev\'e and Gauss Hypergeometric Functions}
\author{A.~V.~Kitaev
\thanks{E-mail: kitaev@pdmi.ras.ru, kitaev@maths.usyd.edu.au}\\
Steklov Mathematical Institute, Fontanka 27, St.Petersburg, 191023, Russia\\
and\\
School of Mathematics and Statistics, University of Sydney,\\ 
Sydney, NSW 2006, Australia}
\date{September 25, 2003}
\maketitle
\begin{abstract} 
We consider an application of Grothendieck's dessins d'enfants to the theory 
of the sixth Painlev\'e and Gauss hypergeometric functions: two classical 
special functions of the isomonodromy type.
It is shown that, higher order transformations and the Schwarz 
table for the Gauss hypergeometric function are closely related 
with some particular Belyi functions. Moreover, we introduce a notion of 
deformation of the dessins d'enfants and show that one dimensional 
deformations are a useful tool for construction of algebraic the sixth 
Painlev\'e functions.
\vspace{24pt}\\
{\bf 2000 Mathematics Subject Classification}: 34M55, 33E17,
33E30.\vspace{24pt}\\
Short title: Dessins d'Enfants and Algebraic Special Functions
\end{abstract} 
\newpage
\setcounter{page}2
\section{Introduction}
 \label{sec:intro}
In this paper we report a further development of the method of 
$RS$-transformations recently introduced \cite{K2} in the theory of
Special Functions of the Isomonodromy Type (SFITs) \cite{K1}.
It was already discussed \cite{K2} that $RS$-transformations are a useful 
tool for solution of many problems in the theory of SFITs, in particular, 
for construction of higher order transformations and special values of these 
generally transcendental functions, say, algebraic values at algebraic 
points. 

Technically the most complicated problem in the method of 
$RS$-transformations is construction of their $R$-parts, i.e., rational 
functions with some special properties. In many cases it is not a priori 
clear whether such rational functions actually exist. 

The progress achieved in this work is based on the observation that the
Belyi functions \cite{B}, which are important in many questions of
algebraic geometry, play also an important role in construction of the 
$RS$-transformations. It is well known that in connection with the theory 
of the Belyi functions Grothendieck \cite{G} suggested the theory of 
``Dessins d'Enfants''. We show that for the theory of SFITs we need not 
only the theory of dessins d'enfants but also a theory of their 
deformations. This theory would help to establish existence of the $R$-parts
of $RS$-transformations. As long as existence of the desired rational 
function is proved it immediately imply existence of the $S$-part of
the corresponding $RS$-transformation. Since construction of the
$S$ part is performed pure algorithmically in terms of the coefficients
of the $R$-part and the original linear ODE, which is subjected
by the $RS$-transformation. Of course, the existence does not tell
how to construct the $R$-part explicitly; it is a separate problem.
However, many theoretical conclusions, e.g., differential equations that 
obey the corresponding SFITs, can be found without their explicit 
constructions. One can write, say, an ansatz for the $RS$-transformation 
with unknown coefficients. The latters can be calculated numerically, or 
studied in some other way. Exact calculation of the monodromy parameters 
of the SFITs constructed via the method of $RS$-transformations also does 
not require explicit expressions, it is enough to know the above mentioned 
ansatz with the numerical values of the coefficients. In fact, the 
existence is helpful even in finding the explicit formulae too: since the
knowledge, that the function $R$ actually exists, stimulates to continuing 
efforts in obtaining these formulae even though a few first attempts 
failed. It is the last comment that was important for the author in
doing this work. It is important to stress in the very beginning, that 
possibly the  most interesting part of this work concerning a relation 
between deformations of the dessins d'enfants and existence of the $R$-parts 
is only conjectured rather than proved.

In this paper, instead of dealing with the general theory of SFITs, we 
continue to consider application of the method of $RS$-transformations
to the theory of two classical one variable SFITs, namely, the sixth
Painlev\'e and Gauss hypergeometric functions. More precisely, we are
interested in explicit constructions of algebraic solutions of the sixth 
Painlev\'e equations \cite{AK2,K2}, a topic that have been attracted 
recently a considerable attention, and also discuss related questions for 
its linear analog, the Gauss hypergeometric function \cite{AK1}.

The main new observation concerning algebraic solutions of the sixth 
Painlev\'e equation which is made in this paper is that a wide class 
of these solutions (possibly all?) can be constructed via a pure algebraic 
procedure without any ``touch'' of differential equations at all! 
The procedure reads as follows:\\
1. Take a proper dessin d'enfant (a bicolour graph);\\
2. Consider its one dimensional deformations (tricolour graphs);\\
3. For each tricolour graph there exists a rational function the $R$-part
of some $RS$-transformation;\\
4. One of the critical points of $R$ is a solution of the sixth Painlev\'e
equation.\\
As is mentioned above we do not have a general proof of the existence
in item 3. In all our examples this existence is obtained via an
explicit construction, which is, of course, technically the most
complicated part of this scheme. At the same time there is a substantial
simplification comparing to the work~\cite{AK2}, as to get explicit
formulae for the algebraic solutions we do not have to find explicitly
$S$-parts of the corresponding $RS$-transformations. Of course, without
the $S$-parts we cannot obtain explicit solutions for the associated 
monodromy problems. But the latters are additional problems which are not
directly related with the original one of finding algebraic solutions for
the sixth Painlev\'e equation.

In \cite{AK2} we began a classification of $RS$-transformations generating
the algebraic sixth Painlev\'e functions. As is explained in the beginning 
of this Introduction the principle problem here is a classification of
$R$-parts of these transformations. It is important to notice that these
$R$-parts can be used in many other $RS$-transformations, say, for SFITs 
related with isomonodromy deformations of matrix ODEs with the matrix 
dimension higher than $2$. Therefore, the problem of classification of
the $R$-parts goes beyond a particular problem related with the 
algebraic sixth Painlev\'e functions. 
The scheme of classification of the $R$-parts that rise out from the
deformation point of view is that it is enough to classify one
dimensional deformations of the dessins d'enfants.  

Our main new result for the Gauss hypergeometric function is an explicit
formulae for the special Belyi function that allows one to construct three
octic transformations which act on finite sets, we call them clusters, 
of transcendental the Gauss hypergeometric functions. Before, the only known
higher order transformations, that act beyond the Schwarz cluster of 
algebraic the Gauss hypergeometric functions, were quadratic and cubic 
transformations, and their compositions.

Now we briefly overview the content of the paper.

In Section~\ref{sec:dessins} we introduce a notion of deformations of the 
dessins. More precisely, we begin with a presentation of the dessins 
as bicolour graphs on the Riemann sphere. Then one dimensional deformations
of the dessins are defined as special tricolour graphs which can be 
obtained by some simple rules from the bicolour ones. After that 
the main statements concerning a relation between the tricolour graphs and
a class of rational functions are formulated as conjectures. Then we establish
a proposition allowing one to calculate algebraic solutions of the sixth 
Painlev\'e equation directly from the latter rational functions. In the 
remaining part of the section we consider some special examples and discuss 
questions important for the theory of the sixth Painlev\'e equation. 

In Section~\ref{sec:P6Platonic} deformations of the dessins
for the Platonic solids are studied. Here and in Section~\ref{sec:dessins} 
we obtained many different algebraic solutions of the
sixth Painlev\'e equations, some of them are, new, the other related with 
the known solutions not in a straightforward way. However, it is only a 
secondary goal of this paper, as well as of the 
papers~\cite{K2,AK2}, to enrich a ``zoology'' of algebraic solutions 
of the sixth Painlev\'e equation. The main purpose is to study different
features of the method of $RS$-transformations and better understand its
place in the theory of SFITs \cite{K1}. Another goal
we achieve in Section~\ref{sec:P6Platonic} is to show that all genus zero 
algebraic solutions of the sixth Painlev\'e equation in the special case 
classified by Dubrovin and Mazzocco~\cite{DM} can be constructed with the 
help of $RS$-transformations. This is aimed towards a check 
of my conjecture~\cite{K2}, that all algebraic solutions for the
sixth Painlev\'e equation can be generated via the method of 
$RS$-transformations and, so-called, the Okamoto transformation 
(see~\cite{KK}). It seems that all algebraic solutions of zero
genus that so far appeared in the literature have been now 
reconstructed by the method of $RS$-transformations or related   
with the ones that are constructed by this method via the certain 
transformations. However, Dubrovin and Mazzocco~\cite{DM} have shown
that there exists one more, genus one, algebraic solution of the sixth 
Painlev\'e equation. At this stage I cannot confirm that this genus one
algebraic solution can be produced in accordance with the above 
conjecture. On the other hand, there are still a few complicated
dessins to be examined to confirm or disprove the conjecture. 

In connection with the conjecture mentioned in the previous paragraph it is 
interesting and instructive to check a closely related, though much simpler 
case, of the Gauss hypergeometric function, especially taking into account 
that a complete classification of the cases when the general solution of the 
Euler equation for the Gauss hypergeometric function is algebraic is known 
due to H. A. Schwarz \cite{SCH}.
Actually, in Section~\ref{sec:Schwarz} we show that the whole Schwarz list
can be generated via the method of $RS$-transformations, whose $R$ parts are
the Belyi functions starting with the simplest Fuchsian ODE with two 
singular points. This possibly, a new constructive point of view allows 
one to find in a straightforward, though in some cases tedious way, 
explicit formulae for all algebraic Gauss hypergeometric functions. We 
call the set of these functions {\it the Schwarz cluster}.  

Section~\ref{sec:non-Schwarz} is a continuation of the previous
work \cite{AK1} devoted to higher order transformations for the Gauss
hypergeometric function. In \cite{AK1} we have found few new higher
order algebraic transformations for the Gauss hypergeometric functions, 
however, all these transformations except quadratic and cubic ones, act 
within the Schwarz cluster. Thus it was an interesting question to 
understand whether there exist transformations of the order higher than 3 
and are not compositions of quadratic and cubic transformations, which 
act on transcendental the Gauss hypergeometric functions. In \cite{AK1} we 
found a numeric construction of an octic transformation which has this
property. Although we were able to find a numerical solution with much more
digits than that indicated in \cite{AK1} and thus it was no any doubt that 
this transformation actually exists, we didn't have a mathematical proof 
of the existence. In this Section an identification of this transformation 
with one of the Belyi functions, immediately gives the desired proof. 
Moreover, by using a better computer the corresponding Belyi function
is calculated explicitly. This makes straightforward an explicit 
construction of, actually, three different octic $RS$-transformations. 
These transformations together with the quadratic and cubic transformations, 
and their inverses define three different clusters of the transcendental 
Gauss hypergeometric functions which are related via algebraic higher 
order transformations, i.e., have the same type of transcendency. We call 
them {\it Octic Clusters} and present them explicitly at the end of 
Section~\ref{sec:non-Schwarz} in the corresponding tables.

Discussing in Sections~\ref{sec:dessins} and~\ref{sec:P6Platonic} different 
questions concerning particular algebraic solutions we make references to 
the quadratic transformations for the sixth Painlev\'e equation. For the 
convenience of the reader in Appendix we give an overview of these 
transformations in the soul of the present work from 
the ``Belyi functions'' angle of view. I hope that even a specialist may 
find this outlook interesting.\\
{\bf Acknowledgement and Comments} After the work was finished and put into 
the web archive I got a letter from P. Boalch, who informed me about two 
very interesting works \cite{Bo,D} that are closely related and 
substantially overlap with this work. I would like to thank him for this 
very important information and a subsequent informal discussion of the
related issues. Below we make necessary comments concerning these papers. 

The solution presented in item 3 (Cross) of 
Subsection~\ref{subsec:icosahedron} was explicitly constructed in the recent 
work by Boalch~\cite{Bo}. He used the method suggested in \cite{DM},
which is very different from the one considered here. 
Moreover, in his work a relation of this solution with the famous Klein's 
quartic algebraic curve in $\mathbb{P}^2$ of genus 3 that has a maximum 
possible number of holomorphic automorphisms is established.

Theorem~\ref{Th:P6solutions} formulated in terms of the $R$-parts rather
than tricolour graphs and modulo explicit formulae for the coefficients 
of the sixth Painlev\'e equation was first established in the work by
Ch.~F.~Doran~\cite{D} (Section 4, Theorem 4.5). Doran's work is based 
essentially on the scalar second order Fuchsian equation and gives a deep 
and general theoretical insight based on many remarkable results known for 
the Belyi functions, Hurwitz spaces, arithmetic Fuchsian groups, etc. 

At the same time Doran does not introduce a new concept of the deformation of 
the dessins like we do, also he does not perform any explicit constructions 
of the algebraic solutions leaving an opportunity for the interested reader
to apply the method by J.-M.~Couveignes. 

Our deformation technique, assuming the validity of Conjectures~\ref{Con:1} 
and \ref{Con:2}, allows one to immediately reproduce the classification 
results of \cite{D} formulated as Corollaries 4.6--4.8 and continue in
a systematic way a production of further ``solvable'' types of the
suitable $R$-parts. Thus, our examples are not just an illustration of the 
classification~\cite{D}: many of them, say, the deformations studied in 
Section~\ref{sec:dessins} or Subsection~\ref{subsec:icosahedron} go
beyond Corollaries~4.6--4.8. I also call attention of the reader to the 
discussion of the renormalization aspect. It is clear that if 
Conjectures~\ref{Con:1} and~\ref{Con:2}) are valid they give an answer on 
the ``inverse problem'', i.e., classification of the types of $R$-parts that 
generate algebraic sixth Painlev\'e's functions (see \cite{AK2} and 
Remark 12 of \cite{D}). Of course, this answer is not absolutely explicit, 
but in principle for any given type a finite number of operations is 
required to check whether there exists the corresponding tricolour graph or 
not and thus give an answer on the inverse problem. I hope that further 
studies will pour more light on these conjectures, so that more explicit 
statements will be available. It is also worth to mention that all $R$-parts 
presented in this work are found by a straightforward though complicated 
method explained in Remark~\ref{Rem:m-parameter}. 

The work by Doran rise a priority question. This question can be separated
into two ones: 1. The idea of using of $RS$-transformations in the
context of SFITs in particular the sixth Painlev\'e equation, and
2. Appearance of the Belyi functions and more generally the Hurwitz curves 
as their $R$-parts.
The method of $RS$-transformations were used by the author~\cite{K4} (1991)
for a construction of the quadratic transformations for the sixth Painlev\'e 
equation. Application of the $RS$-transformations for construction of the 
higher order transformations for SFITs and algebraic SFITs was reported by 
the author at Workshop on Isomonodromic Deformations and Applications in 
Physics, Montreal (Canada), May 1 -- 6, 2000 and published in \cite{K2}.
The works \cite{AK1} and \cite{AK2} were reported at Workshops
in Strasbourg (February, 2001) and Otsu (August, 2001).
We also note that the original scheme of \cite{K2} always uses both
$R$- and $S$- part of the $RS$-transformation. As is clear from this work
it is a more general procedure, than that based on 
Theorem~\ref{Th:P6solutions} (Theorem 4.5 of \cite{D}), some of the known 
solutions we were not able to reproduce without the $S$-parts. Strictly 
speaking this fact does not mean that such solutions cannot be obtained via 
Theorem~\ref{Th:P6solutions} by finding some other suitable $R$-parts, 
however, construction of the $R$-parts is a much more complicated
enterprise than construction of the $S$-parts. 
Note that this is a more general content than that reported in \cite{D}.

At the same time, the fact of the relation of the $R$-parts with the Belyi 
functions and their deformations was noticed by me only in this work, after 
I decided to reproduce via the method of $RS$-transformations the results 
reported in \cite{DM} and also explicitly construct the octic transformation 
for the Gauss hypergeometric function from \cite{AK1}: this requires a more 
detailed study of the corresponding rational functions. So that the second 
fact was first found by Doran and only rediscovered here.
\section{Deformations of Dessins d'Enfants and Algebraic Solutions of the 
Sixth Painlev\'e Equation}
 \label{sec:dessins}
\begin{proposition}
 \label{Prop:necessary_condition}
Let $R:\mathbb{CP}^1\to\mathbb{CP}^1$ be a rational function of degree $n$
with $k\geq3$ critical values. Denote by $k_i$ the number of critical points
corresponding to the $i$-th critical value and put 
\begin{equation}
 \label{eq:parameters}
m=\sum\limits_{i=1}^3k_i-n-2.
\end{equation}
Then $m\geq0$. Moreover, $R$ is the Belyi function with $3$ critical
values iff $m=0$.
\end{proposition}
{\it Proof.}
The Riemann--Hurwitz formula gives
$
\sum\limits_{i=1}^kk_i=(k-2)n+2.
$
Summing up it with Equation~(\ref{eq:parameters}) we arrive at
\begin{equation}
 \label{eq:m-parameter}
m+\sum\limits_{i=4}^kk_i=(k-3)n,
\end{equation}
where we assume that the sum is equal to $0$ if $k<4$. Notice that
$\sum_{i=4}^kk_i\leq(k-3)(n-1)$, therefore, 
Equation~(\ref{eq:m-parameter}) implies $m\geq k-3\geq0$. Moreover,
if $k=3$, then $m=0$ again by virtue of Equation~(\ref{eq:m-parameter}).
\begin{remark}
 \label{Rem:m-parameter}{\rm
The condition $m\geq0$ has a very lucid sense and intuitively
evident: We can assume that the first three critical values are located at 
$0$, $1$, and $\infty$ and define the function $R$ and $R-1$ by two 
rational expressions with indeterminate preimages of $0$, $1$ and $\infty$ 
with the prescribed multiplicities. Writing then the consistency condition 
we arrive at a system of algebraic equations for the indeterminate preimages.
In this setting the condition $m\geq0$ says that the number of equations in 
the system, $n$, should not be greater than the number of unknown parameters, 
$\sum_{i=1}^3k_i-2$. This necessary condition however is not sufficient for 
existence of $R$, see an example in Remark~\ref{rem:counterexample}.
If $m>0$ and the the corresponding function $R$ exists, it may depend on
$m$ parameters. We call, sometimes, such functions $m$ dimensional 
deformation of the Belyi functions.\\ 
I learned from the recent paper~\cite{GT} that this Proposition essentially 
coincides with Silverman's proof of the $abc$-theorem for polynomials. 
A minor difference occurs since in our setting it is natural to count all 
critical points including the point at $\infty$, whilst in the $abc$-setting 
the $\infty$ point is excluded. 
Below in Remark~\ref{Rem:multicolour_graphs} we give also a ``graphical'' 
insight on this Proposition.}
\end{remark}
For a description of rational functions, $z=R(z_1)$, we use a symbol 
which is called their {\it type} and denoted $R(\ldots|\ldots|\ldots)$. 
In the space between two neighbouring vertical lines or the line and one
of the parenthesises, which we call the box, we write a partition of 
$\deg\;R$ into the sum of multiplicities of critical points of the function 
$R(z_1)$ corresponding to one of its critical values in the descending order.
The total number of boxes is normally supposed to coincide with the number 
of critical values of the function $R$. However, in the case $m=1$ which is 
studied in this and the next Sections we do not indicate the fourth 
``evident'' box: $|2+1+\ldots+1|$. At the same time, where convenient we 
include the boxes for non-critical values: $|1+\ldots+1|$.  

Consider on the Riemann sphere bicolour connected graphs; with black and 
white vertices, and faces homeomorphic to a circle. The valencies of the 
black and white vertices are defined in the usual manner. These valencies 
can be equal to any natural number. The vertices of the same colour are not 
connected by edges. We introduce a notion of the {\it black edge}, 
i.e., the path in the graph connecting two black vertices: it contains 
only two black vertices, its end points, and one white vertex. Any two 
black vertices can belong to a few different black edges or cycles. 
Any cycle should contain at least one black vertex. The loops are not
allowed. We also define the {\it black order of a face} as a number 
of the black edges in its boundary. 

With each Belyi function we can associate now a bicolour connected graph
in the following way. The black vertices are preimages of $\infty$ and
the white vertices are preimages of $1$. Their valencies equal to their
respective multiplicities. Each face corresponds to a preimage of $0$, 
the black orders of faces equal to the multiplicities of the corresponding
preimages. Conversely, for each bicolour graph with the properties
stated above, there exists the unique, modulo fractional linear 
transformations of the independent variable, Belyi function whose associated 
bicolour graph is isomorphic to the given one. In case if all valencies of 
the white vertices equal $2$, we do not indicate them at all and, instead of 
a bicolour graph, get a usual planar graph with the black edges and black 
orders of faces coinciding with usual notion of edges and orders of faces. 
Note that after we dismiss the white vertices the loops may appear on the 
latter planar graph.  

Pictures of such bicolour graphs on the Riemann sphere we call 
{\it dessins d'enfants} or just the {\it dessins}. These dessins define on 
the Riemann sphere, so called bipartite maps \cite{S}.

We define now tricolour graphs. It is also connected graphs on the
Riemann sphere with black and white vertices that obey the conditions for 
the bicolour graphs, but also with one blue vertex.   
Any cycle should contain at least one black or blue vertex, any edge 
connecting vertices of different colours. The loops again are not allowed.
The boundary of every face should contain at least one black vertex.    
Valency of the blue vertex equals $4$. Again in case when all the valencies 
of the white vertices equal $2$ we are not indicating them and instead of a 
tricolour graph get a bicolour graph with all black vertices and only one 
blue vertex.  The latter bicolour graph of course, has nothing to do with 
the black-white bicolour graphs introduced in the previous paragraph. In 
particular, the black-blue bicolour graph may contain loops.
For the tricolour graphs we keep the same notions of the black edge and
black order of face, the blue point is not counted in both cases.

Graphically the tricolour graphs can be viewed as obtained from the bicolour 
ones as a result of simple ``deformations'' see examples below and in the
next Section. Therefore we call them the {\it deformation dessins} or very 
often, where there is no cause for a confusion, just the dessins.
\begin{remark}{\rm
 \label{Rem:multicolour_graphs}
At this stage introduction of tricolour graphs with only one blue vertex
looks somewhat artificial; but it is what we need for applications to
the theory of the sixth Painlev\'e equation. If we would think of a
more general rational functions needed for classification of 
$RS$-transformations for multivariable SFITs we have to introduce 
multicolour graphs. Say, via a recurrence procedure by considering one 
dimensional deformations of $n$-colour graphs we arrive to 
$n$- or $n+1$-colour graphs. At each stage we can either add a vertex with
a new or one of the ``old'' colours. We have to admit a coalescence of the 
vertices of the same colours, so that the valencies of the colour vertices 
would be even and $\geq4$. The vertices of the same colour correspond to 
critical points for the same critical value of $R$. The multiplicities of 
these critical points coincides with a half of valencies of the corresponding
colour vertices. Thus the colour vertices correspond to multiple
critical points of $R$. The dimension $m$ of the deformation equals to 
one half of the sum of valencies of the colour vertices minus their total 
number.}
\end{remark}
\begin{conjecture}
 \label{Con:1}
In conditions of Proposition~{\rm\ref{Prop:necessary_condition}} for any 
rational function $z=R(z_1)$ with $m=1$ whose first three critical values 
are $0$, $1$ and $\infty$, there exists a tricolour graph such that:\\
{\rm1}. There is a one-to-one correspondence between its faces, white, and 
black vertices and critical points of the function $R$ for the critical 
values $0$, $1$, and $\infty$, respectively.\\
{\rm2}. Black orders of the faces and valencies of the vertices
coincide with multiplicities of the corresponding critical points.
\end{conjecture}
\begin{remark}{\rm
As the reader will see in Section~\ref{sec:P6Platonic}: with the same
rational function $z=R(z_1)$ with $m=1$ can be associated a few seemingly
different tricolour graphs. The reason is that $R$ actually is the function
of two variables, $R(z_1)=R(z_1,y)$, where $y\in\mathbb{CP}^1$ is a 
parameter. As the function of $y$ $R$ has different branches. The different 
tricolour graphs should be related with these different branches. These also 
should lead to some equivalence relation on the set of the tricolour graphs.
Examples, considered in Items~4 and~5 of Subsection~\ref{subsec:trunccube} 
of Section~\ref{sec:P6Platonic}, show that we cannot just announce all 
tricolour graphs whose associated rational functions have the same type 
equivalent; since these rational functions can be different as the functions 
of $y$. Of course, until this equivalence relation is established in cases
when there are a few different functions $R(z_1,y)$ of $y$ having the same
type as the rational functions of $z_1$ it is complicated to relate them
with the proper tricolour graphs. In the examples of 
Subsection~\ref{subsec:trunccube} referenced above to make such a distinction
we were motivated by a symmetry that exists in one of the examples.
This remark explains why uniqueness for the function-graph correspondence 
in Conjecture~\ref{Con:1} and uniqueness modulo fractional linear 
transformations for the inverse correspondence in the following 
Conjecture~\ref{Con:2} are not stated.}
\end{remark}
\begin{conjecture}
 \label{Con:2}
For any tricolour graph there exists a  function $z=R(z_1,y)$, which is a 
rational function of $z_1\in\mathbb{CP}^1$ with four critical values. 
Three of them: $0$, $1$, and $\infty$, and the corresponding critical 
points are related with the tricolour graph as it is stated in 
Conjecture~{\rm\ref{Con:1}}. The variable $y$ denotes the unique second order 
critical point corresponding to the fourth critical value of $z=R(z_1,y)$.
$R$ is an algebraic function of $y$ of the zero genus.   
\end{conjecture}
\begin{corollary}
 \label{Cor:rational}
In the conditions of Conjecture~{\rm\ref{Con:2}}
There is a representation of the function $R$ as the ratio of coprime
polynomials of $z_1$ such that its coefficients and $y$ allow a simultaneous 
rational parametrization.
\end{corollary}
Consider a rational parametrization of $y=y(s)$ and 
$z=R(z_1,y)\equiv R_1(z_1,s)$ with some parameter $s$ as is stated in
Corollary~\ref{Cor:rational}. We can change the role of the variables $z_1$ 
and $s$, i.e., consider $R_1(z_1,s)$ as the rational function of $s$ and 
treat $z_1$ as an auxiliary parameter. In this case we call $R_1(z_1,s)$ the 
{\it conjugate function} with respect to $R(z_1)$. Making fractional 
linear transformations of $R$ interchanging its critical points we get new 
rational functions of equivalent types. However, their conjugate
functions have, generically, not equivalent types (see examples below).
The reason is that the conjugate functions have dimensions $m\ge1$,
however, instead of $m$ parameters they have only one, $z_1$. 
Critical points of the conjugate functions depend generically on $z_1$. 
However, each function has $m$ critical points independent of $z_1$ we call 
them {\it additional critical points}. The set of the additional critical 
points of all conjugate functions coincides with the set of values of the 
parameter $s$ such that the function $R$ changes its type and therefore
coincides with one of the Belyi functions.

Recall the canonical form of the sixth Painlev\'e equation,
\begin{eqnarray}
 \label{eq:P6}
\frac{d^2y}{dt^2}&=&\frac 12\left(\frac 1y+\frac 1{y-1}+\frac 1{y-t}\right)
\left(\frac{dy}{dt}\right)^2-\left(\frac 1t+\frac 1{t-1}+\frac 1{y-t}\right)
\frac{dy}{dt}\nonumber\\
&+&\frac{y(y-1)(y-t)}{t^2(t-1)^2}\left(\alpha_6+\beta_6\frac t{y^2}+
\gamma_6\frac{t-1}{(y-1)^2}+\delta_6\frac{t(t-1)}{(y-t)^2}\right),
\end{eqnarray}the conjugate functions
where $\alpha_6,\,\beta_6,\,\gamma_6,\,\delta_6\in\mathbb C$ are
parameters. For a convenience of comparison of the results obtained here 
with the ones from the other works we will use also parametrization of the
coefficients in terms of the formal monodromies $\hat\theta_k$:  
$$
\alpha_6=\frac{(\hat\theta_\infty-1)^2}2,\quad
\beta_6=-\frac{{\hat\theta}_0^2}2,\quad\gamma_6=\frac{{\hat\theta}_1^2}2,
\quad\delta_6=\frac{1-{\hat\theta}_t^2}2.
$$

To formulate our main result, showing a relation between the tricolour 
graphs and algebraic solutions of Equation~(\ref{eq:P6}), we recall a 
notion of the $RS$-{\it symbol} for $RS$-trans\-for\-ma\-tions and
introduce some necessary notation.

The $RS$- symbol is designed for a description of $RS$-transformations
for arbitrary SFITs related with the Fuchsian ODEs. Below instead of a
general definition we refer to a more specific situation considered in
this paper. The brief notation for the $RS$-symbols needed here is
$RS_k^2(3)$ for $k=2,3,4$. This notation means that we map a
$2\times2$ matrix Fuchsian ODE with 3 singular point into analogous
ODE but with $k$ singular points.  The extended
notation for the $RS$-symbol, instead of the number $3$ in the parentheses, 
uses three boxes, each one contains  two raws of numbers. In the first raw of
each box there is just a rational number, in the second raw the sum of 
integers representing multiplicities of the preimages of critical values of 
the rational function $R$, the $R$-part of the $RS$-transformation.
It is supposed that three critical values of the function $R$ are located 
at $0$, $1$, and $\infty$. The boxes in the notation of the $RS$-symbol are 
ordered accordingly. For transformations studied in this work we do not have 
a need to indicate other critical values (if any). The whole set of the 
preimages of the critical values: $0$, $1$, and $\infty$ is divided 
(non-uniquely!) on two sets of apparent and non-apparent points: we call them 
{\it apparent and non-apparent sets}, respectively. 

The apparent set is the union of apparent sets of the boxes. 
The apparent set of the $i$-th box, where $i=0$, $1$, $\infty$, consists of 
all points whose multiplicities are divisible by some natural number $\geq2$.
We denote by $n_i\geq2$ the {\bf greatest common divisor} of the apparent 
set of the $i$-th box. In particular, this set can be empty. 
In the last case we formally put  $n_i=1/\theta_i$, where $\theta_i$ is 
a parameter. 
Let a number of the apparent points in the $i$-th box be $N_i\geq0$.
If $N_i\geq1$, then multiplicity of the $j$-th apparent point in the $i$-th 
box can be written as $k^i_jn_i$, where $j=1,\ldots,N_i$ and 
$k^i_j\in\mathbb{N}$.

A critical point of $i$-th critical value is
non-apparent iff its multiplicity is not divisible by $n_i$. The union of
such point is the non-apparent set of $R$. Non-apparent sets for the 
transformations $RS_4^2(3)$, $RS_3^2(3)$ or $RS_2^2(3)$ consist of $4$, $3$, 
or $2$ points, respectively.

In general, we call the $R$-part of some $RS$-transformation with three
or more critical values {\it normalized} iff the set $\{0,1,\infty\}$ is 
a subset of the set of critical values of $R$ and also a subset of
its non-apparent set ( if the non-apparent set consists of only two preimages
it should coincide with the set $\{0,\infty\}$). The rational function 
$R$ with two critical values is normalized, iff the set of its critical
values is a subset of $\{0,1,\infty\}$ and the non-apparent set obey
the same condition as in the previous sentence.    

{\bf Further in this section we consider only 
$RS^2_4(3)$-trans\-for\-ma\-tions}. Suppose that their $R$-part are 
normalized, then the non-apparent set consists of four points:
$0$, $1$, $\infty$, and $t$. Denote their multiplicities as 
$m_0$, $m_1$, $m_\infty$, and $m_t\in\mathbb{N}$.
\begin{definition}
$RS^2_4(3)$-symbol is called special if for $i=0,1$, and $\infty$ the 
rational number in the first raw of the $i$-th box is $1/n_i$.     
\end{definition}
\begin{remark}{\rm
The numbers $1/n_i$ in first raws of the boxes of special $RS$-symbols 
equal to formal monodromy of the $i$-th singular point of the original 
Fuchsian ODEs.}
\end{remark}
\begin{theorem}
 \label{Th:P6solutions}
Let the normalized rational function $z=z(z_1)$ corresponding to a tricolour 
dessin be the $R$-part of some $RS^2_4(3)$-trans\-for\-ma\-tion with a 
special $RS$-symbol. Put $\varepsilon=0$ or $1$ depending on whether
$\sum^3_{i=1}\sum^{N_i}_{j=1}k^i_j$ is even or odd, 
respectively\footnote{The usual convention $\sum^0_1=0$ is assumed.}. 
Then the double critical point $y$ of the function $z(z_1)$, corresponding
to its fourth critical value\footnote{The one different from $0$,$1$, and 
$\infty$.}, considered as the function of the fourth non-apparent critical 
point $t$, is an algebraic solution of the sixth Painlev\'e 
equation~{\rm(\ref{eq:P6})} for the following $\hat\theta$-tuple:
$$
\hat\theta_0=\frac{m_0}{n_{z(0)}},\qquad
\hat\theta_1=\frac{m_1}{n_{z(1)}},\qquad
\hat\theta_t=\frac{m_t}{n_{z(t)}},\qquad
\hat\theta_\infty=\varepsilon+(-1)^\varepsilon\frac{m_\infty}{n_{z(\infty)}}.
$$   
\end{theorem}
{\it Proof.}
We assume that the reader is acquainted with that how the method of
$RS$-transformations is working to produce algebraic solutions of the sixth 
Painlev\'e equation (see \cite{AK2}). A suitable $RS$-transformation can be 
viewed as a composition of some $R$- transformation with a finite number of 
elementary Schlesinger transformations which are successively applied to the 
associated linear $2\times2$ matrix ODE stating with a matrix form of the 
Gauss hypergeometric equation. The key
observation is that for special $RS$-symbols all the elementary
Schlesinger transformations can be chosen to have the same upper-triangular
structure. Thus $y$, the root of the equation $R'(z_1)=0$, is also a root
of $\{21\}$-element of the successively transformed coefficient matrix
of the associated linear ODE at each step of application of the
elementary Schlesinger transformations.
\begin{remark}{\rm
In formulation of Theorem~\ref{Th:P6solutions} we can, of course, instead of 
mentioning of the special $RS$-symbol, formulate all necessary conditions to 
be imposed on the rational function $z(z_1)$ pure graphically,in terms 
of the valencies of the tricolour graph.\\ 
For $\varepsilon=1$ the sign minus in the above formula for 
$\hat\theta_\infty$ is not essential as it does not change the coefficients 
of Equation~(\ref{eq:P6}). We keep it to get smaller absolute values for
$\hat\theta_\infty$.}
\end{remark}
Now we consider some special deformations of the dessins and corresponding
constructions for algebraic solutions of the sixth Painlev\'e equation 
which follows from Theorem~\ref{Th:P6solutions}. 
\begin{remark}{\rm
On all figures throughout the paper we follow the convention that blue 
vertices are indicated similarly to the white ones but with a larger 
diameter than the latters. In case the picture contains only one 
``white vertex'' it is actually the blue one.}
\end{remark}
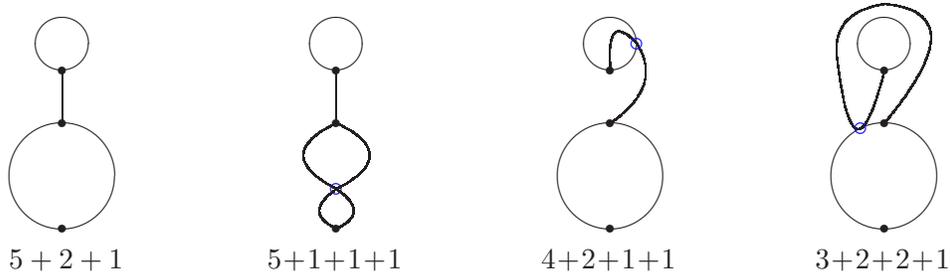
\begin{figure}[ht]
\begin{picture}(40,90)
\put(30,85){\circle{20}}
\put(30,35){\circle{40}}
\put(30,55){\line(0,1){20}}
\put(30,15){\circle*{3}}
\put(30,55){\circle*{3}}
\put(30,75){\circle*{3}}
\put(10,00){$5+2+1$}
\end{picture}
\begin{picture}(40,90)
\put(90,85){\circle{20}}
\put(90,55){\line(0,1){20}}
\put(90,15){\circle*{3}}
\put(90,55){\circle*{3}}
\put(90,75){\circle*{3}}
\put(90,30){\color{blue}\circle{4}}
\qbezier(90,15)(77,21)(90,30)
\qbezier(90,15)(103,21)(90,30)
\qbezier(90,55)(65,43)(90,30)
\qbezier(90,55)(115,43)(90,30)
\put(64,00){$5\!+\!1\!+\!1\!+\!1$}
\end{picture}
\begin{picture}(40,90)
\put(150,85){\circle{20}}
\put(150,35){\circle{40}}
\qbezier(150,75)(148,98)(160,85)
\qbezier(150,55)(170,69)(160,85)
\put(160,85){\color{blue}\circle{4}}
\put(150,15){\circle*{3}}
\put(150,55){\circle*{3}}
\put(150,75){\circle*{3}}
\put(124,00){$4\!+\!2\!+\!1\!+\!1$}
\end{picture}
\begin{picture}(60,100)
\put(210,85){\circle{20}}
\put(210,35){\circle{40}}
\qbezier(192,80)(190,97)(210,100)
\qbezier(210,55)(245,97)(210,100)
\qbezier(192,80)(198,28)(210,75)
\put(201,53){\color{blue}\circle{4}}
\put(210,15){\circle*{3}}
\put(210,55){\circle*{3}}
\put(210,75){\circle*{3}}
\put(184,00){$3\!+\!2\!+\!2\!+\!1$}
\end{picture}
\caption{One parameter ``face'' deformations of the dessin for the Belyi 
function $R(5+2+1|2+2+2+2|3+3+2)$. 
Distributions of the black orders of faces are indicated under the dessins.}
 \label{fig:def521/332}
\end{figure}
On Figure~\ref{fig:def521/332} all one parameter ``face'' deformations for 
the Belyi function $R(5+2+1|2+2+2+2|3+3+2)$ are shown. This Belyi function 
is the dual function to the one that appears later in the proof of 
Proposition~\ref{Prop:non1/2} (see Equation~(\ref{eq:tr6-10})).
Hereafter, we call these and similar deformations of the other dessins
{\it Twist, Cross}, and {\it Join}, respectively. 
Note that two face distributions, namely, $2+2+2+2$ and $3+3+1+1$ which pass 
through the necessary condition of Proposition~\ref{Prop:necessary_condition} 
cannot be realized as the deformations of the first dessin on 
Figure~\ref{fig:def521/332}. However, the latter face 
distribution can be obtained as a face deformation of the dessin for
the Belyi function of the type $R(3+3+2|2+2+2+2|3+3+2)$.
The first one cannot be realized as a deformation of any dessin and thus does 
not define any rational function. There are several other 
seemingly different deformations that lead 
to the same face distributions. In this case, they can be interpreted as 
different branches of the same algebraic solution.
Below we give exact formulae for the deformations of the Belyi function
for all three deformed dessins presented on Figure~\ref{fig:def521/332}
together with the corresponding solutions, $y(t)$, of the sixth Painlev\'e 
equation~(\ref{eq:P6}) calculated via Theorem~\ref{Th:P6solutions}:\\ 
\noindent
{\bf1. Twist}.  
$RS_4^2\left(\!\!\!\!
\begin{tabular}{c|c|c}
$1/5$&$1/2$&$1/3$\\
$5+1+1+1$&$\underbrace{2+\ldots+2}_{4}$&$3+3+2$
\end{tabular}\!\!\!\!
\right)$:
\begin{align*}
&z=\frac{2^{24}s^3(5s^2+118s+5)(s+1)^{10}}
{(3s^3+95s^2+25s+5)^3(5s^3+25s^2+95s+3)^3}\,
\frac{(z_1-1/2-a)^5z_1(z_1-1)(z_1-t)}{(z_1-1/2-c_1)^3(z_1-1/2-c_2)^3},\\
&a=-\frac{(s-1)(s^4+12s^3-410s^2+12s+1)}{128s(s+1)
\sqrt{s(5s^2+118s+5)}},\\
&c_1\equiv c_1(s)=\frac{(5s^6+462s^5+8535s^4+3060s^3+195s^2+30s+1}
{16s(3s^3+95s^2+25s+5)\sqrt{s(5s^2+118s+5)}},\quad c_2=-c_1(1/s),\\
&t=\frac12-\frac{(s-1)(25(s^8+1)+760(s^7+s)+4924(s^6+s^2)+75464(s^5+s^3)
+329174s^4)}{2^{11}s(s+1)^5\sqrt{s(5s^2+118s+5)}},\\
&y=\frac12-\frac{(s-1)(5(s^6+1)+58(s^5+s)+1771(s^4+s^2)+8620s^3)
\sqrt{s(5s^2+118s+5)}}
{8s(s+1)(5s^3+25s^2+95s+3)(3s^3+95s^2+25s+5)},\\
&\qquad\qquad\quad\;\;\;\hat\theta_0=\frac15,\qquad
\hat\theta_1=\frac15,\qquad\hat\theta_t=\frac15,\qquad
\hat\theta_\infty=\frac13.
\end{align*}
It is worth to notice that applying to this solution the Okamoto 
transformation together with the so-called B\"acklund transformations 
we can obtain algebraic solutions of Equation~(\ref{eq:P6}) for the 
following $\hat\theta$-tuples:
$$
\left(\frac1{15},\frac1{15},\frac1{15},\frac7{15}\right)\qquad\&
\qquad\left(\frac4{15},\frac4{15},\frac4{15},\frac2{15}\right).
$$
It is readily seen that by introducing a new variable
$z_2=(z_1-1/2)\sqrt{s(5s^2+118s+5)}$ we get a new function $z(z_2)$
which has the same type as $z(z_1)$ but parametrized by $s$ rationally.
However, to relate our rational function with a solution of the sixth
Painlev\'e equation~(\ref{eq:P6}) we have, as it is explained earlier, to 
{\bf normalize} it by placing, via a fractional linear transformation, three 
of its non-apparent zeroes or poles at $0$, $1$, and $\infty$. In this 
example the normalization procedure leads to the appearance of a genus $1$ 
parametrization. The following Remark~\ref{Rem:symmetric} helps in many cases 
to simplify parametrizations of the $R$-parts of $RS$-transformations 
regardless whether we use them for construction of solutions via 
Theorem~\ref{Th:P6solutions} or with a help of the complete construction
of the corresponding $RS$-transformations including their $S$-parts.
As the reader will see below with this function $z(z_1)$ one can associate
another $RS$-transformation, such that Theorem~\ref{Th:P6solutions}
is not applicable. Thus, one actually have to build up the $S$-part
of the latter $RS$-transformation to find the corresponding solution 
of the sixth Painlev\'e equation.   
\begin{remark}
 \label{Rem:symmetric}{\rm
$S$-parts of the $RS$-trans\-for\-ma\-tions are symmetric functions of the
apparent critical points of their $R$-parts.}
\end{remark}
Thus, we need to know a parametrization of the symmetric functions, 
$c_1+c_2$ and $c_1c_2$, of the apparent critical points $c_1$ and $c_2$ 
rather than their individual parametrization. Of course, the latter leads 
to a simplification of the parametrization. Moreover, it is clear that such 
simplification can theoretically reduce the genus of the parametrization. 
In particular, in item 5 of Subsection~\ref{subsec:trunccube} we have an 
example where it actually happens. Below we show that the simplified 
parametrization can be obtained with the help of the Zhukovski type 
transformation, $s+1/s=2^5s_1+2$. The latter reduces a degree of the 
parametrization by two, however in this case the genus of parametrization 
remains unchanged.
\begin{align*}
&z(z_1)=\frac{(5s_1+4)(8s_1+1)^5}{8(30s_1^3+40s_1^2+10s_1+1)^3}\,
\frac{(z_1-1/2-a)^5z_1(z_1-1)(z_1-t)}{(z_1^2-(1+c_1+c_2)z_1+1/4+c_1c_2
+(c_1+c_2)/2)^3},\\
&a=\!-\frac{(8s_1^2+4s_1-3)s_1}{2\sqrt{s_1(8s_1\!+1)(5s_1\!+4)}},\;\;\!
\frac14+c_1c_2=\!-\frac{320s_1^6+1344s_1^5+1560s_1^4+480s_1^3-60s_1^2+1}
{8(30s_1^3+40s_1^2+10s_1+1)(5s_1+4)},\\
&c_1+c_2=-\frac{(6s_1^2+4s_1-1)(4s_1+5)s_1^2(8s_1+1)}
{(30s_1^3+40s_1^2+10s_1+1)\sqrt{s_1(8s_1+1)(5s_1+4)}},
\end{align*}
\begin{eqnarray*}
&t=\frac12-\frac{(800s_1^4+960s_1^3+312s_1^2+100s_1+15)s_1}
{2(8s_1+1)^2\sqrt{s_1(8s_1+1)(5s_1+4)}},\qquad
y=\frac12-\frac{(40s_1^3+22s_1^2+16s_1+3)\sqrt{s_1(8s_1+1)(5s_1+4)}}
{2(30s_1^3+40s_1^2+10s_1+1)(8s_1+1)},&
\end{eqnarray*}
It is also worth to note that although we have a simpler parametrization of 
the function $z(z_1)$, the first parametrization is still needed in the 
theory of SFITs with several variables, where the critical points $c_1$ and 
$c_2$ are included into the non-apparent set, so that the second 
parametrization just cannot be used for a purpose of construction of 
$RS$-trans\-for\-ma\-tions.

Consider now the associated conjugation functions. It is clear that in this 
case some of them are rational functions on the torus. We consider 
here only conjugate functions rational on the Riemann sphere. To get them 
we have to normalize the function $z(z_2)$ mentioned in the paragraph right 
after the first parametrization of $z(z_1)$. The only way to do it, if we
wish to keep a rational parametrization, is to use for the normalization 
apparent critical points: the zero of order $5$ and poles of the orders $3$, 
$3$, and $2$, together with one non-apparent zero. We call the 
normalization {\it degenerate} if at least one of the critical points $0$, 
$1$, or $\infty$ is apparent. All in all we can arrange $33$ different 
degenerate normalizations of $z(z_2)$ and thus to get $33$ different 
rational conjugation functions. 

For example, the function $z(z_2)$ normalized such that it has a zero of the 
fifth order at $0$, and poles of the third order at $1$ and $\infty$ reads,
\begin{eqnarray}
 \label{eq:5111rat}
&\tilde z=\frac{\tilde z_1^5(\tilde z_1-1+s^2_0)((5s_0+1)^2\tilde z_1^2+
(5s_0+1)(5s_0^3+49s_0^2+115s_0+23)\tilde z_1-24(s_0^2+6s_0+1)(s_0^2+4s_0+1))}
{64(\tilde z_1-1)^3
((3s_0^3+15s_0^2+25s_0+5)\tilde z_1-8(s_0+1)(s_0^2+4s_0+1))^2},&
\end{eqnarray}
This function differs from the one in the beginning of this item, by
a fractional linear transformation of $z_1$ related with the normalization
discussed above: 
$$
\tilde z(\tilde z_1)=z(z_1),\qquad
z_1\equiv M(\tilde z_1)=\frac{\tilde t_{-}-c_0}{\tilde t_{-}-\tilde t_{+}}\,
\frac{\tilde z_1-\tilde t_{+}}{\tilde z_1-c_0},
$$
where
\begin{eqnarray*}
&&\tilde t=1-s_0^2,\qquad
\tilde t_{\pm}=-\frac{24s_0^2}{5s_0+1}-\frac{23+s_0^2}2\pm
\frac{(s_0+5)}2\sqrt{\frac{(5s_0^2+22s_0+5)(s_0+5)}{5s_0+1}},\\
&&c_0=\frac{8(s_0+1)(s_0^2+4s_0+1)}{(3s_0^3+15s_0^2+25s_0+5)},\qquad
s_0=\frac{5-s}{5s-1},
\end{eqnarray*}
and $s$ is exactly the same as in the formulae in the beginning of this item.
The functions $\tilde t_\pm$ are the conjugate roots of the
quadratic polynomial of $z_1$ in the numerator of function~(\ref{eq:5111rat}) 
and $c_0$ is its second order pole.

The function $\tilde z(\tilde z_1)$ has a degenerate normalization. Thus
Theorem~\ref{Th:P6solutions} is not applicable to it. However, in this 
particular case the first order zeroes of $\tilde z(\tilde z_1)$ unexpectedly 
have an ``apparent behaviour'' so that Theorem~\ref{Th:P6solutions} actually
works. Therefore, by putting: $m_0=5$, $m_1=3$, $m_t=1$, $m_\infty=3$, 
$\epsilon=1$, $n_0=1$, and $n_\infty=2$, and formally applying 
Theorem~\ref{Th:P6solutions} to the function $\tilde z(\tilde z_1)$, we
arrive at the conclusion that 
$$
\tilde y=\frac{5(1-s_0^2)(s_0^2+6s_0+1)(s_0^2+4s_0+1)}
{(5s_0+1)(3s_0^3+15s_0^2+25s_0+5)},
$$
being considered as the function of either argument $\tilde t$ or 
$\tilde t_{\pm}$, solves the sixth Painlev\'e equation~(\ref{eq:P6}) for
\begin{equation}
 \label{eq:hattheta5111}
\hat\theta_0=5,\qquad\hat\theta_1=\frac32,\qquad
\hat\theta_t=1,\qquad\hat\theta_\infty=-\frac12.
\end{equation}
Actually, $y(\tilde t)$ and $y(\tilde t_{\pm})$ are just different branches 
of the same genus $0$ algebraic function, so that $\tilde t=\tilde t(s_0)$ 
and $y=y(s_0)$ is its rational parametrization. 

The functions $y(t)$ and $\tilde y(\tilde t)$ are related via the fractional 
linear transformation:
$$
y=M(\tilde y),\qquad t=M(\tilde t).
$$
Of course, the latter transformation can be called ``fractional linear'' only 
conventionally as it parametrized via the same parameter $s_0$ (or $s$) as
$\tilde t$ and $\tilde y$. At this point it is worth to notice that solutions
$y(t)$ and $\tilde y(\tilde t)$ are not related by any of transformations
that act on the set of general solutions of the sixth Painlev\'e equation.  

Moreover, we can treat $\tilde y$ as the function of $\tilde t_1=c_0$ as such
it is a rational function,
$$
\tilde y =\frac{5\tilde t_1(\tilde t_1-2)}{3\tilde t_1-8},
$$
which solves the sixth Painlev\'e equation for the same
$\hat\theta$-tuple (\ref{eq:hattheta5111}). 

Now we consider function~(\ref{eq:5111rat}) as the rational function of
$s_0$ treating $z_1$ as a parameter, i.e., as the conjugate function.
Its type is 
$R(\underbrace{1+\ldots+1}_6|2+2+2|2+2+2)$ and therefore this function has 
four {\it additional} critical points. These points are located at 
$s_0=0$, $\infty$, and $-7/5\pm2\sqrt{6}/5$. We write the corresponding Belyi
functions in terms of the variables $z$ and $z_1$, introduced in the
very beginning of this item, as the values of the function $z(z_1)=z(z_1,s)$
at $s=5$, $1/5$, and $-11/5\pm4\sqrt{6}/5$, respectively. 
The first two Belyi functions have the same type $R(5+1|2+2+2|3+2+1)$, they 
are related with a simple change of argument, $z_1\to1-z_1$; whilst the last 
two functions coincide and their type is 
$R(5+2+1|\underbrace{2+\ldots+2}_4|3+3+2)$:
\begin{eqnarray*}
s=5:&&
z(z_1)=\frac{20(z_1-16/25)^5(z_1-1)}{27z_1(z_1-128/125)^3},\quad
s=\frac15:\quad
z(1-z_1),\\
s=-11/5\pm4\sqrt{6}/5:&&
z=-\frac{5(z_1-27/25)^5z_1^2(z_1-1)}{8(z_1^2+216/125z_1+729/1000)^3}.
\end{eqnarray*}    

For completeness we mention that with this twist one can associate another 
$RS$-transformation, namely,\\
$RS_4^2\left(\!\!\!\!
\begin{tabular}{c|c|c}
$2/5$&$1/2$&$1/3$\\
$5+1+1+1$&$\underbrace{2+\ldots+2}_{4}$&$3+3+2$
\end{tabular}\!\!\!\!
\right)$.
However the corresponding solution of Equation~(\ref{eq:P6}) cannot be 
calculated via Theorem~\ref{Th:P6solutions}.
To find it explicitly one has to construct the $S$ part of the 
transformation like in the work \cite{AK2}. 
Here we only notice that this solution also gain an elliptic
parametrization and corresponds to the following $\hat\theta$-tuple:
\begin{eqnarray*}
\hat\theta_0=\frac25,\qquad
\hat\theta_1=\frac25,&&\hat\theta_t=\frac25,\qquad
\hat\theta_\infty=\frac23.
\end{eqnarray*}
Again the Okamoto transformation together with the B\"acklund ones allow 
one to get algebraic solutions for the following $\hat\theta$-tuples:
$$
\left(\frac2{15},\frac2{15},\frac2{15},\frac{14}{15}\right)\qquad\&
\qquad\left(\frac8{15},\frac8{15},\frac8{15},\frac4{15}\right)
\quad\sim\quad\left(\frac7{15},\frac7{15},\frac7{15},\frac{11}{15}\right).
$$
\noindent
{\bf2. Cross}. 
$RS_4^2\left(\!\!\!\!
\begin{tabular}{c|c|c}
$1/4$&$1/2$&$1/3$\\
$4+2+1+1$&$\underbrace{2+\ldots+2}_{4}$&$3+3+2$
\end{tabular}\!\!\!\!
\right)$:
\begin{eqnarray*}
z\!\!&\!=\!&\!\!-\frac{256s^6}{(3s^2-2)^3(s^2-6)^3}
\frac{(z_1-a)^4(z_1-t)^2z_1(z_1-1)}{(z_1-c_1)^3(z_1-c_2)^3},\quad
c_1=\frac{(s+1)^2(s^2+2s-2)^2}{4s^3(3s^2-2)},\\
a\!\!&\!=\!&\!\!-\frac1{32s^3}(s^2-4s-2)(s^2+2s-2)^2,\qquad
c_2=\frac{(s-2)^2(s^2+2s-2)^2}{16s(s^2-6)},\\
t\!\!&\!=\!&\!\!\frac1{4s^3}(s-2)^2(s+1)^2(s^2+2s-2),\qquad
y(t)=-\frac{(s+1)(s^2+2s-2)(s^2-4s-2)}{2s(s-2)(3s^2-2)(s^2-6)},\\
&&\qquad\qquad\quad\;\;\;\hat\theta_0=\frac14,\qquad
\hat\theta_1=\frac14,\qquad\hat\theta_t=\frac12,\qquad
\hat\theta_\infty=\frac13.
\end{eqnarray*}
The certain transformations as in the previous cases imply now algebraic 
solutions for the following $\hat\theta$-tuples:
$$
\left(\frac1{12},\frac1{12},\frac16,\frac23\right)\qquad\&
\qquad\left(\frac5{12},\frac5{12},\frac16,\frac13\right).
$$
One can associate with this deformation dessin one more $RS$-transformation
with the following symbol
$RS_4^2\left(\!\!\!\!
\begin{tabular}{c|c|c}
$1/2$&$1/2$&$1/2$\\
$4+2+1+1$&$\underbrace{2+\ldots+2}_{4}$&$3+3+2$
\end{tabular}\!\!\!\!
\right)$.
The corresponding explicit solution for $\hat\theta$-tuple
$(1/2,1/2,3/2,-3/2)$ is easy to find by a renormalization
of the function $z(z_1)$ via a fractional linear transformation of $z_1$
analogously to that as it is done in the previous item.\\
The conjugate function is of the type
$R(\underbrace{4+\ldots+4}_6+\underbrace{2+\ldots+2}_6|
\underbrace{2+\ldots+2}_{18}|\underbrace{3+\ldots+3}_{12})
$.
The general function of this type depends on 4 arbitrary parameters.
The conjugate function has four additional critical points: $\pm\sqrt 2$ 
and $\pm i\sqrt 2$. The original function at this values of the parameter 
$s$ coincides with the Belyi functions of the types 
$R(6+1+1|\underbrace{2+\ldots+2}_{4}|3+3+2)$ and 
$R(4+2+1+1|\underbrace{2+\ldots+2}_{4}|6+2)$, respectively:
\begin{eqnarray*}
s=\pm\sqrt{2}:&&z=\frac{(z_1-1/2)^6z_1(z_1-1)}{2(z_1^2-z_1-1/32)^3},\\
s=\sqrt{-2}:&&
z=\frac{(z_1-1/2-5\sqrt{-2}/4)^4(z_1-1/2+11\sqrt{-2}/2)^2z_1(z_1-1)}
{128(z_1-1/2+7\sqrt{-2}/16)^6}.
\end{eqnarray*}
By fractional linear transformations of $z_1$ one can get a few
other non-equivalent conjugate functions, e.g., the one of the type
$
R(\underbrace{2+\ldots+2}_4+\underbrace{1+\ldots+1}_8|
\underbrace{2+\ldots+2}_{8}|6+6+2+2)
$. 
The latter function has $6$ additional critical points located at
$0$, $\infty$, and $\pm1\pm\sqrt{3}$. To all of them corresponds the same 
Belyi function of the type $R(1+1|2|2)$, $z=-4z_1(z_1-1)$.\\   
{\bf3. Join}.
$RS_4^2\left(\!\!\!\!
\begin{tabular}{c|c|c}
$1/3$&$1/2$&$1/3$\\
$3+2+2+1$&$\underbrace{2+\ldots+2}_{4}$&$3+3+2$
\end{tabular}\!\!\!\!
\right)$:
\begin{eqnarray*}
z\!\!&=&\!\!-\frac{(s^2+3)^6}{(s^2+1)^3(s^2+9)^3}
\frac{(z_1-a)^3z_1^2(z_1-1)^2(z_1-t)}{(z_1-c_1)^3(z_1-c_2)^3},\\
a\!\!&=&\!\!\frac{(s^2+6s+3)}{2(s^2+3)},\qquad
c_1=\frac{(s+3)^2}{2(s^2+9)},\qquad
c_2=\frac{(s+1)^2}{2(s^2+1)},
\end{eqnarray*}
\begin{eqnarray*}
t=\frac{(s+1)^2(s+3)^2(s^2-2s+3)}{2(s^2+3)^3},&&
y(t)=\frac{(s+1)(s+3)(s^2-2s+3)(s^2+6s+3)}
{2(s^2+1)(s^2+3)(s^2+9)},\\
\hat\theta_0=\frac23,\qquad\hat\theta_1=\frac23,&&
\hat\theta_t=\frac13,\qquad\hat\theta_\infty=\frac13.
\end{eqnarray*}
By making the fractional linear transformation of $z_1$,
$\tilde z_1=(1-t)z_1/(z_1-t)$, mapping the points $0,1,\infty,t$ to 
$0,1,1-t,\infty$, respectively, we obtain another algebraic solution,
$\tilde y(\tilde t)=(1-t)y(t)/(y(t)-1)$ and  $\tilde t=1-t$. Changing 
the notation $\tilde y$ and $\tilde t$ back to $y$ and $t$ this solution 
can be written as follows,
\begin{eqnarray*}
t=\frac{(s-1)^2(s-3)^2(s^2+2s+3)}{2(s^2+3)^3},&&
y(t)=\frac{(s-1)(s-3)(s^2+6s+3)}
{4s(s^2+3)},\\
\hat\theta_0=\frac23,\qquad\hat\theta_1=\frac23,&&
\hat\theta_t=\frac23,\qquad\hat\theta_\infty=\frac23.
\end{eqnarray*}
For completeness we mention that this dessin generates one more 
$RS$-transformation,
$RS_4^2\left(\!\!\!\!
\begin{tabular}{c|c|c}
$1/2$&$1/2$&$1/2$\\
$3+2+2+1$&$\underbrace{2+\ldots+2}_{4}$&$3+3+2$
\end{tabular}\!\!\!\!
\right)$.
The corresponding solution solves Equation~(\ref{eq:P6}) for 
$\hat\theta$-tuple $(3/2,1/2,3/2,-1/2)$. Its explicit construction 
can be made in a standard way by a renormalizaion of the function $z(z_1)$.

By using the Okamoto, B\"acklund, Quadratic (see Appendix) transformations
and their compositions one can construct from the solutions presented in 
this item many different algebraic solutions. We are not writing here
the corresponding $\hat\theta$-tuples as care must be taken when considering
action of the transformations on $\hat\theta$-tuples: transformations that 
work on general solutions can degenerate on some particular ones and lead 
to ``solutions'' like $y(t)=0$, $y(t)=1$, $y(t)=\infty$. Whether it
happens or not on some intermediate step of a chain of transformations is 
not clear until the actual calculation for the particular solutions is done.\\
The conjugate function is of the type 
$R(3+3+\underbrace{1+\ldots+1}_6|
\underbrace{2+\ldots+2}_{6}|\underbrace{3+\ldots+3}_{4})
$.
It has four additional critical points at $0$, $\infty$, and $\pm\sqrt 3$, and
The original function at this values of the parameter $s$ coincides with the
Belyi functions of the types $R(2+2|2+2|2+2)$ and 
$R(3+2+2+1|\underbrace{2+\ldots+2}_{4}|6+2)$, respectively:
\begin{eqnarray*}
s=0,\,\infty:&&z=-\frac{4z_1^2(z_1-1)^2}{(2z_1-1)^2},\\
s=\pm\sqrt{3}:&&
z=-\frac{27(z_1-1/2\mp\sqrt{3}/2)^3z_1^2(z_1-1)^2 (z_1-1/2\mp5\sqrt{3}/18)}
{64(z_1-1/2\mp\sqrt{3}/4)^6},
\end{eqnarray*}
By making a fractional linear transformation one can get a few other 
nonequivalent conjugate functions; one of them is of the type
$
R(\underbrace{2+\ldots+2}_{8}+\underbrace{1+\ldots+1}_4|
\underbrace{2+\ldots+2}_{10}|6+6+\underbrace{2+\ldots+2}_{4})
$. 
It has six additional critical points at $0$, $\infty$, $1\pm i\sqrt{2}$, 
and $-1\pm i\sqrt{2}$. The Belyi functions corresponding to the last four
critical points, $s=\pm1+\sqrt{-2}$, have the same type
$R(3+3+2|\underbrace{2+\ldots+2}_{4}|3+3+2)$ and by the quadratic
transformation, $\tilde z=(z_1-(1\pm1)/2)^2$, can be reduced to the function 
of the type $R(3+1|2+2|3+1)$, namely, 
$z=-64\tilde z(\tilde z-1)^3/(8\tilde z+1)^3$.
\section{Deformations of Dessins for the Platonic Solids and\\
Algebraic Solutions of the Sixth Painlev\'e Equation} 
 \label{sec:P6Platonic}
The Platonic solids have reach symmetry groups, therefore their dessins can 
be obtained as actions of the certain finite rotation groups on simpler 
dessins called the {\it truncated dessins}. In terms of the corresponding 
Belyi functions it means that these functions are compositions of the Belyi 
functions for the truncated dessins and a monomial $z=z_1^n$ with some 
integer $n$. We refer to the interesting paper \cite{MZ} for details. 
As is shown in this section and the following one these irreducible 
truncated dessins are very important in the theory of algebraic the Sixth 
Painlev\'e and Gauss hypergeometric functions.

In the case when the $\hat\theta$-tuple is of the form 
$(0,0,0,\hat\theta_\infty$), where $\hat\theta_\infty\in\mathbb C$ is a 
parameter,  algebraic solutions for the sixth Painlev\'e equations were 
classified by Dubrovin and Mazzocco \cite{DM}. They found five cases, 
namely: $\theta_\infty=-1/2$, so-called, Tetrahedron solution, 
$\theta_\infty=-2/3$ - Cube solution, another solution for 
$\theta_\infty=-2/3$ - Great Dodecahedron solution, 
$\theta_\infty=-4/5$ - Icosahedron solution, and 
$\theta_\infty=-2/5$ - Great Icosahedron solution. 
They also gave a rational parametrization for all solutions, except the Great 
Dodecahedron one. They mention that the last one is an algebraic function of 
genus $1$ and produced via computer simulations incredibly long algebraic 
equation for this solution, which can be found in the preprint version of 
their work\footnote{http://arXiv.org/abs/math/9806056}. 
Here we show that all algebraic solutions of zero genus mentioned above
can be constructed via the method of $RS$-transformations whose 
$R$-parts are obtained as deformations of the dessins for the Platonic solids.
Here we introduce also different kinds of vertex deformations of the dessins
which we call {\it Splits}. Instead of the splits one can always consider 
the face deformations of the dual dessins.
\subsection{Deformations of Folded Truncated Cube (Tetrahedron Solution)}
 \label{subsec:tetrahedron}
Below we define, the folded truncated cube dessin. This a very simple dessin 
has four one dimensional deformations: twist, join, B-, and W-splits.
However, the first three of them generate equivalent 
$RS$-trans\-for\-ma\-tions: the B-split is dual to the twist and the latter 
defines the same face distribution as the join. The last two dessins can be 
continuously transformed one into another, when the blue vertex passes 
through the white one. Thus they correspond to two different branches of the 
same function $z(z_1)$ (as the function of the deformation parameter $s$) 
and define the same algebraic solution of the sixth Painlev\'e equation.  
So there are two nonequivalent tricolour dessins. With each one can be 
associated one (seed) algebraic solution. Both solutions were constructed
via the method of $RS$-trans\-for\-ma\-tions in Section3 of \cite{AK2},
where, of course, their deformation nature was not discussed.\vspace{3pt}\\
\parbox{195pt}
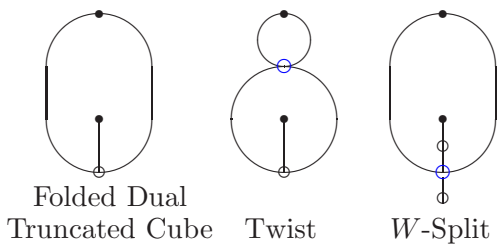
{
\begin{picture}(60,95)
\put(35,55){\oval(40,60)}
\put(35,25){\line(0,1){20}}
\put(35,25){\circle{4}}
\put(35,45){\circle*{3}}
\put(35,85){\circle*{3}}
\put(10,12){Folded Dual} 
\put(00,00){Truncated Cube} 
\put(105,45){\oval(40,40)}
\put(105,75){\circle{20}}
\put(105,25){\line(0,1){20}}
\put(105,65){\color{blue}\circle{5}}
\put(105,25){\circle{4}}
\put(105,45){\circle*{3}}
\put(105,85){\circle*{3}}
\put(90,12){}
\put(90,00){Twist}
\put(165,55){\oval(40,60)}
\put(165,25){\line(0,1){20}}
\put(165,13){\line(0,1){10}}
\put(165,25){\color{blue}\circle{5}}
\put(165,45){\circle*{3}}
\put(165,85){\circle*{3}}
\put(165,35){\circle{4}}
\put(165,15){\circle{4}} 
\put(145,00){$W$-Split}
\end{picture}}
\parbox{7.9cm}{
The Belyi function corresponding to the truncated 
cube is of the type $R(4+\!1\!+\!1|2+2+2|3+3)$ (see \cite{MZ}).
Interchanging the colours of white and black vertices we go to the dual 
dessin. The folding procedure for the latter dessin is the
transition to the depicted dessin based on the
following decomposition of the types of the Belyi functions, 
$R(4+1+1|3+3|2+2+2)=R(\underset{\wedge}{2}+1|3|2+\underset{\wedge}{1})
\circ R(2|2)$.}
The symbol of the $RS$-trans\-for\-ma\-tion associated with the 
$W$-split reads\\
$RS_4^2\left(\!\!
\begin{tabular}{c|c|c}
$\theta_0$&1/2&1/2\\
2+1&2+1&2+1
\end{tabular}\!\!
\right)$. The complete list of formulae related with this 
$RS$-trans\-for\-ma\-tion including the corresponding solution
can be found in \cite{AK2} Subsection~3.1.1.
Below we consider only the twist, which generates Tetrahedron Solution
of \cite{DM}, treating it with the help of  Theorem~\ref{Th:P6solutions}. 
Corresponding $RS$-symbol is 
$RS_4^2\left(\!\!
\begin{tabular}{c|c|c}
$\theta_0$&1/3&1/2\\
1+1+1&3&2+1
\end{tabular}\!\!
\right)$. 
\begin{eqnarray*}
z=(1-s^2)^3\frac{z_1(z_1-1)(z_1-t)}{\left((1-s^3)z_1-1\right)^2},&&
z-1=\frac{\left((1-s^2)z_1-1\right)^3}{\left((1-s^3)z_1-1\right)^2},\\
t=\frac{(2s+1)}{(1-s)(s+1)^3},&&y(t)=\frac{(2s+1)}{(s+1)(s^2+s+1)},\\
\hat\theta_0=\hat\theta_1=\hat\theta_t\in\mathbb{C},&&\hat\theta_\infty=1/2.
\end{eqnarray*}
Since $\hat\theta_0=\hat\theta_1=\hat\theta_t$ arbitrary and $y(t)$ does not 
depend on them we arrive at the following algebraic equation for $y(t)$:
$$
\frac{t}{y^2}-\frac{(t-1)}{(y-1)^2}+\frac{t(t-1)}{(y-t)^2}=0;
$$
Actually, the solution found and called in \cite{DM} Tetrahedron solution
is more complicated and corresponds to a more special $\hat\theta$-tuple,
$\hat\theta_0=\hat\theta_1=\hat\theta_t=0$, $\hat\theta_\infty=-1/2$.
The latter solution is related with the one presented here via a B\"acklund
transformation and the associated Fuchsian linear ODE has the same monodromy 
group. See details in Remark 2 of Subsection 3.1.2 of 
\cite{AK2}\footnote{There is a misprint in formula for $\rho$ given in this
Subsection, denominator of this formula should be squared.}.
\subsection{Deformations of Truncated Tetrahedron (Cube Solution)}
 \label{subsec:cube}
There are two deformations of Truncated Tetrahedron with each one we can 
associate one solution. Both solutions were obtained via the method of
$RS$-transformations in \cite{K1,AK2}. However here we provide the first one
with a simpler parametrization than that given in the cited works and 
discuss application of the quadratic transformations to the second solution.
The quadratic transformations are interesting in the context of finding
algebraic solutions with nontrivial genus, as their application often leads 
to the solutions with elliptic parametrization.\bigskip\\
\parbox{270pt}
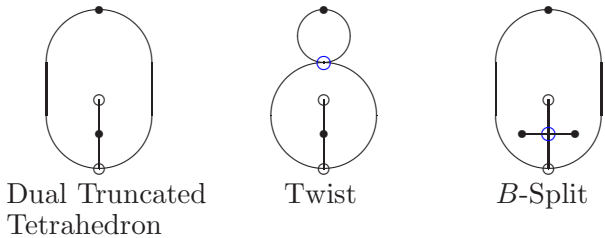
{
\begin{picture}(60,95)
\put(35,55){\oval(40,60)}
\put(35,25){\line(0,1){26}}
\put(35,51){\circle{4}}
\put(35,25){\circle{4}}
\put(35,38){\circle*{3}}
\put(35,85){\circle*{3}}
\put(00,12){Dual Truncated} 
\put(00,00){Tetrahedron} 
\put(120,45){\oval(40,40)}
\put(120,75){\circle{20}}
\put(120,25){\line(0,1){26}}
\put(120,65){\color{blue}\circle{5}}
\put(120,25){\circle{4}}
\put(120,51){\circle{4}}
\put(120,38){\circle*{3}}
\put(120,85){\circle*{3}}
\put(105,12){Twist}
\put(205,55){\oval(40,60)}
\put(205,25){\line(0,1){26}}
\put(205,51){\circle{4}}
\put(205,25){\circle{4}}
\put(205,38){\color{blue}\circle{5}}
\put(195,38){\line(1,0){20}}
\put(205,85){\circle*{3}}
\put(195,38){\circle*{3}}
\put(215,38){\circle*{3}}
\put(185,12){$B$-Split} 
\put(170,00){}
\end{picture}}
\parbox{150pt}{ The Belyi function for the Dual Truncated Tetrahedron reads,
$z=\frac{64z_1(z_1-1)^3}{(8z_1^2+20z_1-1)^2}$ and
$z-1=-\frac{(8z_1+1)^3}{(8z_1^2+20z_1-1)^2}$.}\bigskip
\newline
{\bf1. Twist} (Cube Solution):
$RS_4^2\left(\!\!
\begin{tabular}{c|c|c}
$\theta_0$&1/3&1/2\\
2+1+1&3+1&2+2
\end{tabular}\!\!
\right)$.
\begin{eqnarray*}
&z=\frac{64(z_1-t)^2z_1(z_1-1)}{(8z_1^2-2(s+1)(s^2-s+4)z_1+(1+s)^3)^2},\quad
z-1=\frac{(4sz_1-(s+1)^2)^3}{(8z_1^2-2(s+1)(s^2-s+4)z_1+(1+s)^3)^2},&
\end{eqnarray*}
\begin{eqnarray*}
t=\frac14(2-s)(s+1)^2,&&y(t)=\frac1{2s}(s+1)(2-s),\\
\hat\theta_0=\hat\theta_1=\frac12\hat\theta_t\in\mathbb{C},&&
\hat\theta_\infty=\frac23.
\end{eqnarray*}
Similar to the Subsection~\ref{subsec:tetrahedron} no efforts required to 
derive an algebraic equation for the function $y(t)$\footnote{In \cite{K2} 
there is a misprint in the sign before the last term in this equation},
$$
\frac{t}{y^2}-\frac{(t-1)}{(y-1)^2}+\frac{4t(t-1)}{(y-t)^2}=0;
$$ 
Concerning the relation of this solution to Cube solution of
\cite{DM} one can make analogous remark as at the end of
Subsection~\ref{subsec:tetrahedron}: the latter Cube solution satisfies 
the sixth Painlev\'e equation only for 
$\hat\theta_0=\hat\theta_1=\frac12\hat\theta_t=0$
$\theta=-2/3$ and related to the one constructed here via a 
B\"acklund transformation.
\newline
{\bf2. $B$-Split} (Cube Solution?):
$RS_4^2\left(\!\!
\begin{tabular}{c|c|c}
1/3&1/3&1/2\\
3+1&3+1&2+1+1
\end{tabular}\!\!
\right)$.
\begin{eqnarray*}
z=\frac{\rho(z_1-a)^3z_1}{(z_1-c)^2(z_1-1)},\quad
z-1=\frac{\rho(z_1-b)^3(z_1-t)}{(z_1-c)^2(z_1-1)},&&
\rho=\frac{(2s+1)^3}{(3s+1)^2(1-3s^2)^2},
\end{eqnarray*}
\vspace{-7mm}
\begin{eqnarray*}
a=\frac{(1-3s^2)}{(2s+1)}(3s^2+2s+1),&& b=1-3s^2,\quad
c=\frac{(1-3s^2)}{(3s+1)}(3s^2+3s+1),\\
t=\frac{(1-3s^2)}{(2s+1)^3}(3s^2+3s+1)^2,&&
y(t)=\frac{(3s^2+2s+1)(3s^2+3s+1)}{(2s+1)(3s+1)},\\
\hat\theta_0=\frac13,\qquad\hat\theta_1=\frac12,&&
\hat\theta_t=\frac13,\qquad\hat\theta_\infty=\frac12.
\end{eqnarray*}
We can further apply the quadratic transformation given in Example~3 of 
Appendix~\ref{app:A}. Put in this Example $\hat\theta_t^0=2/3$ and 
$\hat\theta^0_\infty=0$. Then we see that the $\hat\theta$-tuple of  
Example~3 coincides with the $\hat\theta$-tuple for the solution obtained 
above, so that we can make {\it inverse} quadratic transformation of 
the above solution. Redenoting variables $t_0$ and $y_0(t_0)$ back as 
$t$ $y(t)$ we arrive at the following solution:
\begin{eqnarray*}
&t=\frac12+\frac{(3s+1)(s+1)(3s^3(3s+2)+(s+1)^2)}
{4\left(\sqrt{s(2s+1)(3s+2)}\right)^3},\quad
y(t)=\frac12-\frac{9(s+1)^4-2(3s+2)^2}{6(s+1)(3s+1)\sqrt{s(2s+1)(3s+2)}},&\\
&\hat\theta_0=0,\quad
\hat\theta_1=0,\quad
\hat\theta_t=\frac23,\quad
\hat\theta_\infty=0.&
\end{eqnarray*}
We see that the resulted solution has an elliptic parametrization.
However, the study of its Puiseux expansions at the singular points
$t=0,1,\infty$
shows that they looks very similar to the ones for Cube solution.
So, most probably, this solution can be rationally parametrized and
via a B\"aklund transformation interchanging 
$(0,0,\theta_t,0)\longmapsto(0,0,0,\theta_t)$, mapped to Cube solution.

So here we have described a mechanism of appearance of elliptic 
parametrization of algebraic solutions via an application of the quadratic 
transformations. This mechanism is seemingly different from the normalization
one explained in Section~\ref{sec:dessins}. 
At the moment it is not clear whether such mechanisms could
really produce an elliptic algebraic solution.
As it is explained in Appendix~\ref{app:A} the quadratic transformations are 
also generated by $RS$-transformations. However, to get the complete list of 
the quadratic transformations by applying the method of $RS$-transformations 
to the $2\times2$ matrix Fuchsian system with four singularities in
Jimbo-Miwa parametrization, we need the Okamoto transformation, which can be 
treated as a reparametrization of the Fuchsian system. So that we can 
use the method of $RS$-transformations to the Fuchsian system in the
Okamoto parametrization also for production of the algebraic solutions.   
In the latter setting we do not need to separately apply any quadratic
transformations: we just apply the same $RS$-transformations to two
copies of the Fuchsian system in the different parametrizations. 
Therefore, if one can actually produce via the method 
of $RS$-transformations an algebraic solution with a nontrivial genus, then 
its appearance can be ``explained'' as a special case of  the first 
normalization mechanism of Section~\ref{sec:dessins}.

Since throughout the paper we often reference the Okamoto transformation,
it is probably worth to notice, that this transformation does not change $t$ 
and is rational in $y(t)$, $y'(t)$, and $t$. Thus it cannot change a genus 
of our algebraic solutions. 
\subsection{Deformations of Truncated Cube}
 \label{subsec:trunccube}
On the pictures below all nonequivalent one dimensional deformations of 
Truncated Cube are presented. On the first four dessins the white vertices
are not indicated. It is interesting that here we have two nonequivalent
$W$-Splits which define the same face distributions.\\
\parbox{270pt}
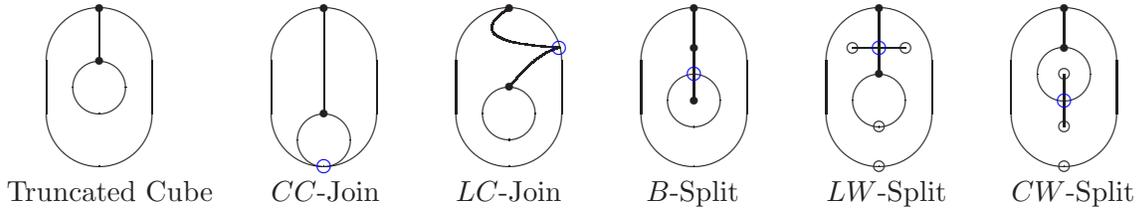
{
\begin{picture}(60,95)
\put(35,55){\oval(40,60)}
\put(35,55){\oval(20,20)}
\put(35,65){\line(0,1){20}}
\put(35,65){\circle*{3}}
\put(35,85){\circle*{3}}
\put(00,12){Truncated Cube} 
\put(00,00){} 
\put(120,55){\oval(40,60)}
\put(120,35){\oval(20,20)}
\put(120,45){\line(0,1){40}}
\put(120,45){\circle*{3}}
\put(120,85){\circle*{3}}
\put(120,25){\color{blue}\circle{5}}
\put(100,12){$CC$-Join}
\put(190,55){\oval(40,60)}
\put(190,45){\oval(20,20)}
\qbezier(209,70)(170,72)(190,85)
\qbezier(209,70)(200,68)(190,55)
\put(190,55){\circle*{3}}
\put(190,85){\circle*{3}}
\put(209,70){\color{blue}\circle{5}}
\put(170,12){$LC$-Join} 
\put(260,55){\oval(40,60)}
\put(260,50){\oval(20,20)}
\put(260,50){\line(0,1){35}}
\put(260,70){\circle*{3}}
\put(260,50){\circle*{3}}
\put(260,85){\circle*{3}}
\put(260,60){\color{blue}\circle{5}}
\put(242,12){$B$-Split}
\put(330,55){\oval(40,60)}
\put(330,50){\oval(20,20)}
\put(330,60){\line(0,1){25}}
\put(320,70){\line(1,0){20}}
\put(330,60){\circle*{3}}
\put(320,70){\circle{4}}
\put(340,70){\circle{4}}
\put(330,40){\circle{4}}
\put(330,25){\circle{4}}
\put(330,85){\circle*{3}}
\put(330,70){\color{blue}\circle{5}}
\put(310,12){$LW$-Split}
\put(400,55){\oval(40,60)}
\put(400,60){\oval(20,20)}
\put(400,70){\line(0,1){15}}
\put(400,40){\line(0,1){20}}
\put(400,70){\circle*{3}}
\put(400,60){\circle{4}}
\put(400,85){\circle*{3}}
\put(400,25){\circle{4}}
\put(400,40){\circle{4}}
\put(400,50){\color{blue}\circle{5}}
\put(380,12){$CW$-Split}
\end{picture}}
\newline
{\bf1. $CC$-Join}. 
The function $z(z_1)$, as the function of $y$, has two branches. 
The dessin corresponding to the second branch has the edge going from the 
upper B-vertex around the inner circle, joining with itself, and finally 
connecting with the B-vertex on the inner circle. We can associate two 
$RS$-transformations with this dessin:\\
$RS_4^2\left(\!\!
\begin{tabular}{c|c|c}
1/2&1/2&$\theta$\\
2+2+1+1&2+2+2&3+3
\end{tabular}\!\!
\right)$ and
$RS_4^2\left(\!\!
\begin{tabular}{c|c|c}
$\theta$&1/2&1/3\\
2+2+1+1&2+2+2&3+3
\end{tabular}\!\!
\right)$.
Notice that the $R$-type of these transformations can be presented as
a composition of $R$-types of the degrees $3$ and $2$, 
$R(2+2+1+1|2+2+2|3+3)=R(2+1|2+\underset{\wedge}{1}|3)\circ R(2)$. This
results in that both transformations generate exactly the same 
solution as
$RS_4^2\left(\!\!
\begin{tabular}{c|c|c}
$\theta_0$&1/2&$\theta_\infty$\\
1+1&2&1+1
\end{tabular}\!\!
\right)$, but formally they give different and more restricted 
$\hat\theta$-tuple than the latter. The reader can find this solution 
in \cite{AK2} Section~2.\\
{\bf2. $LC$-Join}. The function $z(z_1)$ as the function of $y$ has three 
branches. Two other branches correspond to two Crosses: one of the inner 
circle another of the external. Corresponding $RS$-transformation is
$RS_4^2\left(\!\!
\begin{tabular}{c|c|c}
$\theta_0$&1/2&1/3\\
3+1+1+1&2+2+2&3+3
\end{tabular}\!\!
\right)$.
\begin{align}
&z=-\frac{4(p_1-p_3)p_1^4p_3^4}{3^6(4s+1)^9(10s+1)^9}\,
\frac{z_1(z_1-1)(z_1-t)}{(z_1-c_1)^3(z_1-c_2)^3},\quad
z-1=-\frac{(z_1^3+b_2z_1^2+b_1z_1+b_0)^2}{(z_1-c_1)^3(z_1-c_2)^3},
\nonumber\\
&b_0=\frac{s^3(11s+2)^3p_3^3}{(4s+1)^6(10s+1)^6},\quad
b_1=\frac{p_3^2(p_1^4+p_2^4)}{3^3(4s+1)^6(10s+1)^6},\quad
b_2=-\frac{p_3(p_1^2+p_2^2)}{3(4s+1)^3(10s+1)^3},\nonumber\\
&c_1=\frac{(11s+2)^2p_3}{3^2(4s+1)^3(10s+1)},\quad
c_2=\frac{3^2s^2p_3}{(4s+1)(10s+1)^3},\label{eq:3111itd}\\
&\qquad
t=\frac{3^3s^3(11s+2)^3p_3}{(p_1-p_3)(4s+1)^3(10s+1)^3},\qquad\qquad
y(t)=\frac{3^2s^2(11s+2)^2}{(p_1-p_3)(4s+1)(10s+1)},\nonumber\\
&\qquad\qquad\qquad\quad
\hat\theta_0=\hat\theta_1=\hat\theta_t=\theta_0\in\mathbb{C},\qquad\qquad
\hat\theta_\infty=3\theta_0+1.\nonumber
\end{align}
where $p_k$ $k=1,2,3$ are the quadratic polynomials:
$$
p_1=73s^2+20s+1,\quad
p_2=37s^2+11s+1,\quad
p_3=47s^2+22s+2.
$$ 
They satisfy the following relations:
$p_1+p_2=(11s+2)(10s+1)$, $p_1-p_2=9s(4s+1)$,
$p_3+p_2=3(4s+1)(7s+1)$, $p_3-p_2=(s+1)(10s+1)$.  
Since $\theta_0$ is arbitrary we find the following algebraic equation
for the function $y(t)$,
$$
\frac{t}{y^2}-\frac{t-1}{(y-1)^2}+\frac{t(t-1)}{(y-t)^2}=9.
$$
{\bf3. $B$-Split}. This dessin defines a function $z(z_1)$ of the 
following type $R(4+1+1|2+2+2|3+2+1)$. It is interesting to note that
as the function of $y$ it has three branches. 
\parbox{280pt}{Only two of them
can be obtained as $B$-splits of Truncated Cube. The third branch is
defined by the twist of another Belyi function of the following type
$R(4+2|2+2+2|3+2+1)$, which is not related with the Platonic Solids
(see the picture).
With the $B$-split we can associate four (seed) $RS$-transformations,}
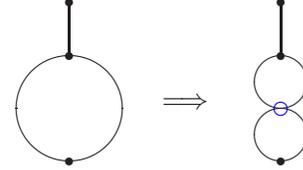
\parbox{120pt}{
\begin{picture}(120,70)
\put(40,20){\oval(40,40)}
\put(40,40){\line(0,1){20}}
\put(40,00){\circle*{3}}
\put(40,40){\circle*{3}}
\put(40,60){\circle*{3}}
\put(75,20){$\Longrightarrow$}
\put(120,10){\circle{20}}
\put(120,30){\circle{20}}
\put(120,40){\line(0,1){20}}
\put(120,60){\circle*{3}}
\put(120,40){\circle*{3}}
\put(120,00){\circle*{3}}
\put(120,20){\color{blue}\circle{5}}
\end{picture}}\vspace{6pt}
\begin{align}
\label{eq:RS411222itd}
&RS_4^2\left(\!\!
\begin{tabular}{c|c|c}
1/4&1/2&1/3\\
4+1+1&2+2+2&3+2+1
\end{tabular}\!\!
\right),\quad\phantom{and}
&RS_4^2\left(\!\!
\begin{tabular}{c|c|c}
1/4&1/2&1/2\\
4+1+1&2+2+2&3+2+1
\end{tabular}\!\!
\right),\\
&RS_4^2\left(\!\!
\begin{tabular}{c|c|c}
1/2&1/2&1/3\\
4+1+1&2+2+2&3+2+1
\end{tabular}\!\!
\right),\quad {\rm and}
&RS_4^2\left(\!\!
\begin{tabular}{c|c|c}
1/2&1/2&1/2\\
4+1+1&2+2+2&3+2+1
\end{tabular}\!\!
\right).\nonumber
\end{align}
To find solutions corresponding to the last two $RS$-trans\-for\-ma\-tions 
one has to construct also their Schlesinger part. We will not consider it 
here, just note that corresponding algebraic solutions has genus $0$ 
and solves Equation~(\ref{eq:P6}) for the following $\hat\theta$-tuples:
$$ 
\left(\frac12,\frac12,\frac13,\frac13\right),\qquad\&\qquad
\left(\frac12,\frac12,\frac12,\frac12\right).
$$
We call attention that to the solution corresponding to the first of
these $\hat\theta$-tuples one can apply the quadratic 
transformation~\ref{eq:quadratic-kit} (see Appendix~\ref{app:A}) to get 
a solution for the following $\hat\theta$-tuple, $(1/3,1/3,1/3,1/3)$. 
The further application of the Okamoto transformation produces a solution 
for $(0,0,0,2/3)$. This construction should be examined in view of
Great Dodecahedron solution.  

The solution associated with the first two $RS$-trans\-for\-ma\-tion can be 
found with the help of Theorem~\ref{Th:P6solutions}:
\begin{eqnarray*}
z=\rho\,\frac{(z_1-a)^4z_1(z_1-1)}{(z_1-c)^3(z_1-t)},&&
z-1=\rho\,\frac{(z_1^3+b_2z_1^2+b_1z_1+b_0)^2}{(z_1-c)^3(z_1-t)},
\quad c=\frac{(s+1)^4}{2^3s(s^2+1)},\\
\rho=-\frac{2^6s^6}{(3s^2-1)(s^2+1)^3},&&
a=\frac{(3s-1)(s+1)^3}{2^4s^3},\quad
b_0=-\frac{(2s-1)(s+1)^8}{2^9s^6},\\
b_2=-\frac{(3s^4+12s^3+6s^2-1)}{2^3s^3},&&
b_1=\frac{(15s^4+36s^3-22s^2+4s-1)(s+1)^4}{2^8s^6},
\end{eqnarray*}
\begin{eqnarray*}
t=\frac{(s+1)^4(2s-1)^2}{8s^3(3s^2-1)},&&
y(t)=\frac{(3s-1)(2s-1)(s+1)^3}{4s(3s^2-1)(s^2+1)},\\
\hat\theta_0=\frac14,\qquad
\hat\theta_1=\frac14,&&
\hat\theta_t=\frac13,\qquad
\hat\theta_\infty=\frac13.
\end{eqnarray*}  
Starting from this solution and using  
the quadratic transformations and Okamoto transformation 
(see  Appendix~\ref{app:A}) one can obtain algebraic solutions for the
following $\hat\theta$-tuples:
$$
\left(\frac12,\frac12,\frac13,\frac14\right),\qquad
\left(0,0,\frac1{12},\frac7{12}\right),\qquad
\left(\frac1{24},\frac1{24},\frac5{24},\frac{19}{24}\right),\qquad
\left(\frac7{24},\frac7{24},\frac{11}{24},\frac{13}{24}\right).
$$ 
The solution corresponding to the second 
$RS$-trans\-for\-ma\-tion~(\ref{eq:RS411222itd})
is a renormalization of the function $z(z_1)$:
$$
\tilde z=z(M^{-1}(\tilde z_1)),\qquad
\tilde z_1=M(z_1)\equiv\frac{(1-c)z_1}{z_1-c};
\qquad
\tilde t=M(t),\qquad\tilde y=M(y).
$$ 
Redenoting back $\tilde t\to t$ and $\tilde y\to y$ we get:
\begin{eqnarray*}
t=-\frac{(2s-1)^2}{8s},&&
y(t)=\frac{(s-1)(3s-1)(2s-1)}{12s(s^2+1)},\\
\hat\theta_0=\frac14,\qquad
\hat\theta_1=\frac14,&&
\hat\theta_t=\frac12,\qquad
\hat\theta_\infty=-\frac12.
\end{eqnarray*} 
To the general solution for the latter $\hat\theta$-tuple all kinds of the 
known transformations are applicable and generate solutions for the 
following $\hat\theta$-tuples:
$$
\left(\frac14,\frac14,\frac14,\frac14\right)\!,\;\;
\left(0,0,\frac12,0\right)\!,\;\;
\left(\frac12,\frac12,\frac12,\frac14\right)\!,\;\;
\left(0,0,\frac14,\frac34\right)\!,\;\;
\left(\frac18,\frac18,\frac18,\frac78\right)\!,\;\;
\left(\frac38,\frac38,\frac38,\frac58\right)\!.
$$ 
Care should be taken to check whether application of some transformations
will not degenerate this particular solution, i.e., do not map them to 
$0$, $1$, or $\infty$.\\  
{\bf4. $LW$-Split}. 
The function $z(z_1)$ as the function of $y$ has two branches.
Another branch corresponds to the twist of the dessin which is a 
multiplication of Folded Truncated Cube\vspace{2pt}\\
\parbox{280pt}{and the segment corresponding to the following composition 
of the types for the associated Belyi functions,
$R(4+2|2+2+1+1|3+3)=
R(\underset{\wedge}{2}+\underset{\wedge}{1}|2+1|3)\circ R(2|2)$ 
(see the picture).
The type of the resulted dessin is also a composition of the types of 
degrees $3$ and $2$, 
$R(4+1+1|2+2+1+1|3+3)=R(\underset{\wedge}2+1|2+1|3)\circ R(2)$. 
Therefore $LW$-Split dessin generate exactly
the same solution as the one in item 1 
($CC$-Join)} 
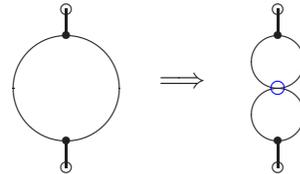
\parbox{120pt}{
\begin{picture}(120,55)
\put(40,30){\oval(40,40)}
\put(40,50){\line(0,1){10}}
\put(40,00){\line(0,1){10}}
\put(40,10){\circle*{3}}
\put(40,50){\circle*{3}}
\put(40,60){\circle{4}}
\put(40,00){\circle{4}}
\put(75,30){$\Longrightarrow$}
\put(120,20){\circle{20}}
\put(120,40){\circle{20}}
\put(120,50){\line(0,1){10}}
\put(120,00){\line(0,1){10}}
\put(120,60){\circle{4}}
\put(120,50){\circle*{3}}
\put(120,10){\circle*{3}}
\put(120,00){\circle{4}}
\put(120,30){\color{blue}\circle{5}}
\end{picture}}\vspace{6pt}
\newline 
and we refer again to Section~2 of \cite{AK2} .
The $RS$-transformation associated with the $LW$-split is
$RS_4^2\left(\!\!
\begin{tabular}{c|c|c}
1/4&1/2&1/3\\
4+1+1&2+2+1+1&3+3
\end{tabular}\!\!
\right)$.
Therefore, this solution corresponds to the following $\hat\theta$-tuple 
$\left(1/4,1/4,1/2,1/2\right)$. However, since the same solution can be 
produced via a simpler $RS$-trans\-for\-ma\-tion, it solves 
Equation~(\ref{eq:P6}) for a more general $\hat\theta$-tuple. It is actually 
interesting to compare this solution with the one for $CW$-Split in the next 
item as the latter has exactly the same $RS$-symbol. Therefore, we provide 
below the details of this $RS$-trans\-for\-ma\-tion. Our solution differs 
from the one cited in \cite{AK2} by fractional linear transformations of $y$ 
and $t$.
\begin{align*}
&z=\frac{(z_1-a)^4z_1(z_1-t)}{(z_1-c_1)^3(z_1-c_2)^3},
&&z-1=-\frac{3(s^2+4s+1)^2}{s^2(s+2)^2}\,
\frac{(z_1^2+b_1z_1+b_0)^2(z_1-1)}{(z_1-c_1)^3(z_1-c_2)^3},\\
&a=\frac{(s^2+4s+1)}{(s+2)s},
&&c_1=\frac{(s^2+4s+1)}{(s+2)^2},\qquad
c_2=-\frac{(s^2+4s+1)}{3s^2},\\
&b_0=\frac{(s^2+4s+1)^2}{9s^2(s+2)^2},
&&b_1=-\frac{2(5s^2+2s-4)(s^2+4s+1)}{9s^2(s+2)^2},
\end{align*}
\begin{align*}
&t=\frac{(s^2-1)(s^2+4s+1)}{s^2(s+2)^2},&
&y(t)=\frac{(s^2-1)}{(s+2)s},\\
&\hat\theta_0=\hat\theta_t\in\mathbb{C},&
&\hat\theta_1=\hat\theta_\infty-1\in\mathbb{C}.
\end{align*}
The algebraic equation for $y(t)$ is very simple, $(y-1)^2=1-t$.\vspace{3pt}
\newline\vspace{3pt}
\parbox{317pt}{{\bf5. $CW$-Split}.
Exactly the same deformation as $CW$-Split, as well as a few others which 
are related with different branches of the same function $z(z_1)$ can be
also obtained as a cross and joins of the dessin shown on the picture.
$RS$-symbol of the transformation associated with  $CW$-Split reads exactly 
the same as for $LW$-Split,}
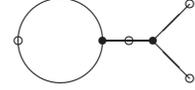
\parbox{100pt}{
\begin{picture}(30,35)
 \label{fig:9}
\put(30,15){\circle{30}}
\put(45,15){\line(1,0){20}}
\put(65,15){\line(1,1){14}}
\put(65,15){\line(1,-1){14}}
\put(14,15){\circle{3}}
\put(56,15){\circle{3}}
\put(79,29){\circle{3}}
\put(65,15){\circle*{3}}
\put(46,15){\circle*{3}}
\put(79,01){\circle{3}}
\end{picture}}
$RS_4^2\left(\!\!
\begin{tabular}{c|c|c}
1/4&1/2&1/3\\
4+1+1&2+2+1+1&3+3
\end{tabular}\!\!
\right)$.
However its $R$-part the function $z(z_1)$ as the function of $y$ is 
different! It is important to mention that now we keep the denominator
of function $z(z_1)$ in the non-factorized form. If we factorize it, like 
in the previous item, then only an elliptic parametrization of the function 
$z(z_1)$ is possible.
\begin{align*}
&z=\frac{(z_1-a)^4z_1(z_1-t)}{(z_1^2+c_1z_1+c_0)^3},
&&z-1=\frac{3(s^2-2)^2(s^2-4s-2)^2}{16(s+1)^3(s-1)^3}\,
\frac{(z_1^2+b_1z_1+b_0)^2(z_1-1)}{(z_1^2+c_1z_1+c_0)^3},
\end{align*}
\begin{eqnarray*}
&a=\frac{(s^2-2s+2)(s^2-2)}{4(s-1)^3},\qquad
c_1=-\frac{((s^2+2)^2-4s(s-2)^2)(s^2-2)}{12(s-1)^3(s+1)},\qquad
c_0=\frac{(s^2-2)^2(s^2+2)^2}{48(s+1)(s-1)^5},&
\end{eqnarray*}
\begin{align*}
&b_0=-\frac{(s^2-2)^2(s^2+2)^3}{144(s^2-4s-2)(s-1)^6},
&&b_1=-\frac{(s^2-2)(s^5-6s^4+10s^3-32s^2+12s-20)}{18(s^2-4s-2)(s-1)^3},
\end{align*}
\begin{align*}
&t=\frac{(s^2-2)(s^2+2)^3}{16(s+1)^3(s-1)^3},&
&y(t)=-\frac{(s^2-2s+2)(s^2+2)^2}{4(s+1)(s^2-4s-2)(s-1)^2},\\
&\hat\theta_0=\hat\theta_t=\frac14,&
&\hat\theta_1=\hat\theta_\infty=\frac12.
\end{align*}
We can further apply to this solution a quadratic transformation, the 
inverse to the one in Example 3 of Appendix~\ref{app:A}. As the result we get
a new solution again denoted as $y(t)$:
\begin{eqnarray*}
&t=\frac12+\frac{(s^8-28s^6+96s^4-112s^2+16)}
{16s\left(\sqrt{(1-s^2)(s^2-4)}\right)^3},\qquad
y(t)=\frac12+\frac{s(7s^4-44s^2+28)}
{2((s^2-2)^2-16s^2)\sqrt{(1-s^2)(s^2-4)}},&
\\
&\hat\theta_0=0,\qquad\hat\theta_1=0,\qquad\hat\theta_t=\frac12,\qquad
\hat\theta_\infty=0.&
\end{eqnarray*}
Due to the obvious symmetry we can use the Zhukovski transformation to 
parametrize this solution rationally:
\begin{eqnarray*}
\frac{s^2}2+\frac2{s^2}=\frac{5-s_1^2}2:&&
t=\frac{(s_1-1)(s_1+3)^3}{16s_1^3},\qquad
y(t)=\frac{(s_1+1)(s_1+3)^2}{2s_1(15+s_1^2)}.
\end{eqnarray*}
This solution can be mapped further to produce Tetrahedron solution 
considered in Subsection~\ref{subsec:tetrahedron}. The $\hat\theta$-tuple 
for $LW$-Split solution of the previous item are such that one can also 
try to apply to it the same quadratic transformation, however it is 
exactly one of those exceptional solutions for which this transformation 
fails: it maps this solution into the critical values of the sixth Painlev\'e 
equation~(\ref{eq:P6}), depending on the choice of the branches to $0$ or $1$.
\subsection{Deformations of Truncated Dodecahedron (Icosahedron Solutions)}
 \label{subsec:icosahedron}
\begin{figure}[ht]
\begin{picture}(60,110)
\put(40,65){\oval(60,80)}
\put(40,65){\oval(40,60)}
\put(40,55){\circle{20}}
\put(40,35){\line(0,1){10}}
\put(40,95){\line(0,1){10}}
\put(40,35){\circle*{3}}
\put(40,45){\circle*{3}}
\put(40,95){\circle*{3}}
\put(40,105){\circle*{3}}
\put(20,10){Truncated} 
\put(10,00){Dodecahedron}
\end{picture}
\begin{picture}(60,100)
\put(90,65){\oval(60,80)}
\put(90,55){\oval(40,40)}
\put(90,85){\circle{20}}
\put(90,55){\circle{20}}
\put(90,35){\line(0,1){10}}
\put(90,95){\line(0,1){10}}
\put(90,75){\color{blue}\circle{4}}
\put(90,45){\circle*{3}}
\put(90,35){\circle*{3}}
\put(90,95){\circle*{3}}
\put(90,105){\circle*{3}}
\put(75,00){Twist}
\end{picture}
\begin{picture}(60,100)
\put(140,65){\oval(60,80)}
\put(140,65){\oval(40,60)}
\put(150,65){\circle{20}}
\put(160,65){\color{blue}\circle{4}}
\qbezier(140,35)(138,43)(150,55)
\put(140,95){\line(0,1){10}}
\put(150,55){\circle*{3}}
\put(140,35){\circle*{3}}
\put(140,95){\circle*{3}}
\put(140,105){\circle*{3}}
\put(125,00){Join}
\end{picture}
\begin{picture}(60,110)
\put(190,65){\oval(60,80)}
\put(200,85){\circle{20}}
\put(190,50){\oval(40,30)}
\qbezier(190,105)(170,80)(190,65)
\qbezier(190,35)(210,65)(200,75)
\put(203,63){\color{blue}\circle{4}}
\put(190,35){\circle*{3}}
\put(200,75){\circle*{3}}
\put(190,65){\circle*{3}}
\put(190,105){\circle*{3}}
\put(175,00){Cross} 
\end{picture}
\caption{Some face deformations of the dessin for the Belyi 
function of Truncated Dodecahedron 
$R(5+5+1+1|2+2+2+2+2+2|3+3+3+3)$.}
\label{fig:truncdodecahedron}
\end{figure}
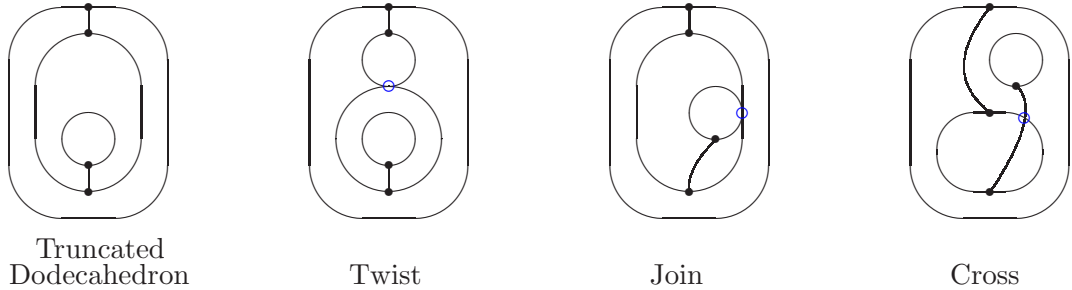

Truncated Icosahedron and its dual solid Truncate Dodecahedron and 
their Belyi functions were introduced by Magot and Zvonkin~\cite{MZ} 
in connection with their study of the Belyi functions of the 
Archimedian solids. Here we consider only face deformations of the 
corresponding dessins.
They are indicated on Figure~\ref{fig:truncdodecahedron}.
\newline
\parbox{337pt}{Actually the last dessin on 
Figure~\ref{fig:truncdodecahedron} is a join of the dessin for the 
Belyi function of the type 
$R(8+2+1+1|\underbrace{2+\ldots+2}_6|\underbrace{3+\ldots+3}_4)$
(see the picture), rather than the cross of Truncated Dodecahedron. However, 
we consider it here since it is interesting to get a solution for the 
$\hat\theta$-tuple proportional to $1/7$.}
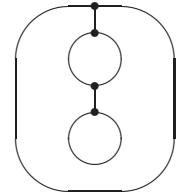
\parbox{80pt}{
\begin{picture}(60,80)
\put(40,40){\oval(60,70)}
\put(40,55){\circle{20}}
\put(40,25){\circle{20}}
\put(40,35){\line(0,1){10}}
\put(40,65){\line(0,1){10}}
\put(40,35){\circle*{3}}
\put(40,45){\circle*{3}}
\put(40,65){\circle*{3}}
\put(40,75){\circle*{3}}
\end{picture}}
\newline
{\bf1. Twist}. 
This deformation produces both, Icosahedron and Great Icosahedron
solutions of \cite{DM} and also one more solution which is related with 
Tetrahedron solution of \cite{DM}.
\begin{align*}
&z=\frac{3^3(s^2+3)^5(s^2-5)^5(s^2+4s-1)^5(s^2-4s-1)}
{2^{10}(s+3)^{12}(s-1)^{20}}\,
\frac{(z_1-a)^5z_1(z_1-1)(z_1-t)}{(z_1^4+c_3z_1^3+c_2z_1^2+c_1z_1+c_0)^3},\\
&a=\frac{2^4(s^2-5)}{(s-1)(s^2+3)(s+3)^3},\quad
c_3=-\frac{(s^2-5)(s^6+4s^5-3s^4-8s^3+115s^2-60s+15)}{(s-1)^5(s+3)^3},\\
&c_2=\frac{(s^2-5)^2}{2^4(s-1)^{10}(s+3)^6}
\left(s^{12}+8s^{11}+10s^{10}-40s^9+1135s^8+3408s^7-10036s^6\right.\\
&\left.-14160s^5+71055s^4-78040s^3+39050s^2-9480s+1185\right),\quad
c_0=\frac{2^8s^2(s^2-5)^4}{(s+3)^{10}(s-1)^{10}},\\
&c_1=\frac{8(s^2-5)^3(15-105s+525s^2-705s^3+107s^4+461s^5+183s^6+29s^7+2s^8)}
{(s-1)^{10}(s+3)^9},
\end{align*}
Consider the following $RS$-symbol:
$RS_4^2\left(\!\!
\begin{tabular}{c|c|c}
1/5&1/2&1/3\\
$5+4+1+1+1$&$\underbrace{2+\ldots+2}_6$&$\underbrace{3+\ldots+3}_4$
\end{tabular}\!\!
\right)$, the corresponding solution found via 
Theorem~\ref{Th:P6solutions} is as follows:
\begin{eqnarray*}
t=-\frac{2^8s^3(s^2-5)}{(s-1)^5(s+3)^3(s^2-4s-1)},&&
y(t)=-\frac{2^6s^2}{(s-1)(s+3)(s^2+3)(s^2-4s-1)},\\
\hat\theta_0=\hat\theta_1\!\!&\!\!=\!\!&\!\!\hat\theta_t=
\hat\theta_\infty=\frac15.
\end{eqnarray*}
This solution can be mapped by the Okamoto transformation 
(Appendix~\ref{app:A}) to the one for the $\hat\theta$-tuple
$\left(0,0,0,2/5\right)$ and then, by a B\"acklund transformation,
to Great Icosahedron solution of \cite{DM} corresponding to the following 
$\hat\theta$-tuple, $\left(0,0,0,-2/5\right)$.

Another $RS$-transformation that is associated with this deformation is\\
$RS_4^2\left(\!\!
\begin{tabular}{c|c|c}
2/5&1/2&1/3\\
$5+4+1+1+1$&$\underbrace{2+\ldots+2}_6$&$\underbrace{3+\ldots+3}_4$
\end{tabular}\!\!
\right)$. 
To find explicit formulae for the corresponding solution of
the sixth Painlev\'e equation one has to find the $S$-part of this
$RS$-transformation similar to the examples of \cite{AK2}. The corresponding
$\hat\theta$-tuple reads $\left(2/5,2/5,2/5,2/5\right)$. By means of 
the same transformations as for the previous solution, this one
can be mapped to Icosahedron Solution of \cite{DM} corresponding to the 
following $\hat\theta$-tuple, $\left(0,0,0,-4/5\right)$.

We can construct one more $RS$-transformation by rearranging the 
normalization of the function $z(z_1)$, i.e., with the help of the
function $\tilde z(\tilde z_1)=z\left(M^{-1}(\tilde z_1)\right)$, where 
the fractional linear transformation $M$ is defined as follows,
$\tilde z_1\equiv M(z_1)=\frac{(1-a)z_1}{z_1-a}$. Then the
function $\tilde y(\tilde t)$, where $\tilde t=M(t)$ and $\tilde y=M(y)$ is
a new solution of the sixth Painlev\'e equation (\ref{eq:P6}).
Omitting the sign $\sim$ in the notation of the new solution we get:
\begin{eqnarray*}
t=\frac{2^4s^3}{(s+3)^3(s-1)},&&
y(t)=\frac{4(s^2+4s-1)s^2}{5(s+3)(s^2+3)(s-1)},\\
\hat\theta_0=\hat\theta_1=\hat\theta_t=\frac14,&&\hat\theta_\infty=-\frac14.
\end{eqnarray*} 
This solution by the Okamoto transformation can be mapped to the one
for the $\hat\theta$-tuple $\left(0,0,0,1/2\right)$ and the latter,
according to \cite{DM}, has to coincide with Tetrahedron Solution.
\newline
{\bf2. Join}. This dessin also produces a few algebraic solutions of the 
Sixth Painlev\'e equation.\\
The first transformation is generated by the symbol\\ 
$RS_4^2\left(\!\!\!\!
\begin{tabular}{c|c|c}
1/5&1/2&1/3\\
$5+3+2+1+1$&$\underbrace{2+\ldots+2}_6$&$\underbrace{3+\ldots+3}_4$
\end{tabular}\!\!\!\!
\right)$ whose $R$-part reads as follows,
\begin{align*}
&z=-\frac{2^53^3(s^2-5)^5}{(s+3)^9(s-2)^6}\,
\frac{(z_1-a)^5(z_1-1)^2z_1(z_1-t)}{(z_1^4+c_3z_1^3+c_2z_1^2+c_1z_1+c_0)^3},
\quad a=\frac{(s-1)(s^2-5)}{2^3(s-2)^2},\\
&c_0=\frac{s^2(s^2-5)^4}{2^4(s+3)^4(s-2)^6},\quad
c_1=-\frac{(8s^4-35s^3+65s^2-75s+45)(s^2-5)^3}{4(s+3)^4(s-2)^6},\\
&c_2=\frac{5(s-1)(2s^3+2s^2-3s-9)(s^2-5)^2}{2(s-2)^4(s+3)^4},\quad
c_3=-\frac{2(s^2-5)(2s^3+5s^2-15)}{(s+3)^3(s-2)^2}.
\end{align*} 
Then Theorem~\ref{Th:P6solutions} provide the corresponding solution, 
\begin{eqnarray}
t=\frac{2s^3(s^2-5)}{(s-2)^2(s+3)^3},&&
y(t)=\frac{s^2(s-1)}{3(s-2)(s+3)},\label{eq:dodecjoin}\\
\hat\theta_0=\frac15,\qquad\hat\theta_1=\frac25,&&
\hat\theta_t=\frac15,\qquad\hat\theta_\infty=\frac25.\nonumber
\end{eqnarray}
With this $R$-part one can associate another $RS$-transformation
by choosing $2/5$ instead of $1/5$ in the first box of the  
$RS$-symbol written above.
However, to find the corresponding solution of the sixth 
Painlev\'e equation, one has to construct explicitly the $S$-part 
of this transformation. The corresponding $\hat\theta$-tuple of 
the latter solution is $\left(6/5,4/5,2/5,2/5\right)$. By means of 
the certain transformations this solution can be mapped to the one 
corresponding to exactly the same $\hat\theta$-tuple as the 
the first solution $y(t)$. A natural question, whether these solutions 
are different, requires a further investigation. It is worth to 
notice that by the Okamoto transformation solution~(\ref{eq:dodecjoin}) 
can be mapped to a solution for the following $\hat\theta$-tuple 
$\left(0,0,1/5,3/5\right)$. The latter solution by the quadratic 
transformation given in item 3 of Appendix~\ref{app:A} can be further 
transformed to solutions for the following $\hat\theta$-tuples:
$$
\left(\frac1{10},-\frac15,\frac1{10},\frac45\right)\quad {\rm and}\quad 
\left(\frac3{10},-\frac25,\frac3{10},\frac35\right).
$$

Another transformation associated with this join is defined by the symbol\\
$RS_4^2\left(\!\!\!\!
\begin{tabular}{c|c|c}
1/3&1/2&1/3\\
$5+3+2+1+1$&$\underbrace{2+\ldots+2}_6$&$\underbrace{3+\ldots+3}_4$
\end{tabular}\!\!\!\!
\right)$. Its $R$-part, $\tilde z(\tilde z_1)$, is obtained from the 
$R$-part of the first $RS$-transformation by the following renormalizing 
fractional linear transformation,
$$
\tilde z(\tilde z_1)=z\left(M^{-1}(\tilde z_1)\right),\quad
\tilde z_1=M(z_1)\equiv\frac{(1-a)z_1}{z_1-a}.
$$
The solution is $\tilde y(\tilde t)$, where $\tilde t=M(t)$ and 
$\tilde y=M(y(t))$, with $t$ and $y(t)$ are defined by 
Equations~(\ref{eq:dodecjoin}). As usual omitting the sign $\sim$, we have 
the following explicit parametric form for $\tilde y(\tilde t)$:
\begin{eqnarray*}
t=\frac{2s^3}{(s+1)(s-2)^2},&&
y(t)=-\frac{(s-3)s^2}{5(s+1)(s-2)},\\
\hat\theta_0=\frac13,\qquad\hat\theta_1=\frac23,&&
\hat\theta_t=\frac13,\qquad\hat\theta_\infty=-\frac23.
\end{eqnarray*}

Finally, there is one more normalization of the function $z(z_1)$ which
produces one more $RS$-transformation,
$RS_4^2\left(\!\!\!\!
\begin{tabular}{c|c|c}
1/2&1/2&1/3\\
$5+3+2+1+1$&$\underbrace{2+\ldots+2}_6$&$\underbrace{3+\ldots+3}_4$
\end{tabular}\!\!\!\!
\right)$. Normalizing transformation reads,
$$
\tilde z(\tilde z_1)=z\left(M^{-1}(\tilde z_1)\right),\quad
\tilde z_1=M(z_1)\equiv\frac{z_1}{z_1-a}.
$$
The corresponding solution of the sixth Painlev\'e equation~(\ref{eq:P6})
is given in terms of the latter fractional linear transformation $M$ and the 
solution $y(t)$ (see Equations~(\ref{eq:dodecjoin})) exactly by the same 
formulae as the ones for $\tilde y(\tilde t)$ from the previous paragraph,
\begin{eqnarray*}
t=-\frac{16s^3}{(s+1)(s-3)^3},&&
y=\frac{8s^2(s-2)}{5(s+1)(s-3)^2},\\
\hat\theta_0=\frac12,\qquad\hat\theta_1=\frac32,&&
\hat\theta_t=\frac12,\qquad\hat\theta_\infty=-\frac32.
\end{eqnarray*}
{\bf3. Cross}. With this dessin we can associate four $RS$-transformations.
The first one is
$RS_4^2\left(\!\!\!\!
\begin{tabular}{c|c|c}
1/7&1/2&1/3\\
$7+2+1+1+1$&$\underbrace{2+\ldots+2}_6$&$\underbrace{3+\ldots+3}_4$
\end{tabular}\!\!\!\!
\right)$. Its $R$-part and the corresponding algebraic solution are as 
follows:
\begin{align*}
&z=-\frac{3^3(3s^2+1)^7}{(7s^2+1)^4}\,
\frac{(z_1-a)^7z_1(z_1-1)(z_1-t)}{z_1^4+c_3z_1^3+c_2z_1^2+c_1z_1+c_0)^3},\\
&a=-\frac{(s-1)(2s^2+s+1)^2}{2(3s^2+1)},\quad
c_3=\frac{2(s-1)(57s^6+57s^5+71s^4+22s^3+15s^2+s+1)}{(7s^2+1)^2},\\
&c_2=-\frac{(42s^6-42s^5+161s^4-44s^3+16s^2-10s+5)(2s^2+s+1)^2}{2(7s^2+1)^2},
\\
&c_1=-\frac{7(s-1)(3s^4-3s^3+6s^2-3s+1)(2s^2+s+1)^4}{2(7s^2+1)^2},\quad
c_0=\frac{(3s-1)^2(2s^2+s+1)^6}{16(7s^2+1)^2},
\end{align*}
\begin{eqnarray}
t=-\frac{(2s^2+s+1)^2(3s-1)^3}{2(7s^2+1)^2},&&
y(t)=-\frac{(s-1)(2s^2+s+1)(3s-1)^2}{2(7s^2+1)(3s^2+1)},
\label{eq:dodeccross}\\
\hat\theta_0=\hat\theta_1=\hat\theta_t=\frac17,&&\hat\theta_\infty=\frac57.
\nonumber
\end{eqnarray}

Two other $RS$-transformations are based on the same $R$-part but the first
box of their $RS$-symbol contains $2/7$ and $3/7$ instead of $1/7$. 
To get explicit formulae for the corresponding solutions
one has to construct the $S$-part of these transformations. Here we only 
mention that these solutions corresponds to the following $\hat\theta$-tuples:
$$
\left(\frac27,\frac27,\frac27,\frac47\right)\qquad {\rm and}\qquad
\left(\frac37,\frac37,\frac37,\frac17\right).
$$
\begin{remark}{\rm
It is interesting to note that if we apply the Okamoto and certain other
transformations to solution~(\ref{eq:dodeccross}) we can get algebraic 
solutions exactly for both $\hat\theta$-tuples written above! It should be 
however checked whether solutions constructed via the Okamoto transformation 
and $RS$-transformations coincide?}
\end{remark}

The fourth $RS$-transformation is
$RS_4^2\left(\!\!\!\!
\begin{tabular}{c|c|c}
1/2&1/2&1/3\\
$7+2+1+1+1$&$\underbrace{2+\ldots+2}_6$&$\underbrace{3+\ldots+3}_4$
\end{tabular}\!\!\!\!
\right)$. Its $R$-part is the following renormalization $\tilde z(\tilde z_1)$
of the function $z(z_1)$:
$$
\tilde z(\tilde z_1)=z\left(M^{-1}(\tilde z_1)\right),\quad
\tilde z_1=M(z_1)\equiv\frac{(1-a)z_1}{z_1-a}.
$$
The new solution $\tilde y(\tilde t)$ is given by the usual formulae:
$\tilde t=M(t)$, $\tilde y=M(y(t))$, where $t$ and $y(t)$ are given by
Equations~(\ref{eq:dodeccross}). Explicit form for $\tilde y(\tilde t)$
(with the omitted sign $\sim$) is:
\begin{eqnarray*}
t=\frac{(3s-1)^3(s+1)}{16s},&&
y(t)=-\frac{(2s^2-s+1)(3s-1)^2}{14s(3s^2+1)},\\
\hat\theta_0=\hat\theta_1=\hat\theta_t=\frac12,&&\hat\theta_\infty=-\frac52.
\end{eqnarray*}
\section{The Schwarz Cluster}
 \label{sec:Schwarz}
Recall the Euler equation for the Gauss hypergeometric function, 
\begin{equation}
 \label{eq:euler}
z(1-z)\frac{d^2u}{dz^2}+(c-(a+b+1)z)\frac{du}{dz}-abu=0.
\end{equation} 
Define the parameters,
$$
\lambda=1-c,\quad\mu=b-a,\quad\nu=c-a-b,
$$
and introduce in the set of triples $(\lambda,\mu,\nu)$ an equivalence 
relation: two triples are called equivalent iff one can be transformed
into another by a permutation and transformation of the form
$$
\lambda\to l\pm\lambda,\quad\mu\to m\pm\mu,\quad\nu\to n\pm\nu,
$$
where the integers $l,m,n$ are such that $l+m+n$ is an even number.
If a triple is equivalent to the one that obey the
following relation $\lambda+\mu+\nu=1$, than it is called degenerate.
In the degenerate case one independent solution is an elementary function,
the other can be expressed in terms of elementary and the incomplete Beta
functions via an application of a finite sequence of simple transformations. 

H.~A.~Schwarz {\cite{SCH} proved that in the non-degenerate case the general 
solution of Equation~(\ref{eq:euler}) is an algebraic function iff 
the corresponding parameters $\lambda,\mu,\nu$ can be reduced to one of 
the 15 cases listed in the following table.
\par
\begin{table}[ht]
 \caption{The Schwarz List}
  \label{tab:1}
\begin{center}
{\renewcommand{\arraystretch}{0}%
\begin{tabular}{|c|ccc||c|ccc||c|ccc||c|ccc||c|ccc|}
\hline
\strut$N$&$\lambda$&$\mu$&$\nu$&$N$&$\lambda$&$\mu$&$\nu$&
$N$&$\lambda$&$\mu$&$\nu$&$N$&$\lambda$&$\mu$&$\nu$&
$N$&$\lambda$&$\mu$&$\nu$\\
\hline
$1\rule{0pt}{12pt}$&$\frac12$&$\frac12$&$\frac pn$&$4$&$\frac12$&$\frac13$&
$\frac14$&
$7$&$\frac25$&$\frac13$&$\frac13$&$10$&$\frac35$&$\frac13$&$\frac15$&
$13$&$\frac45$&$\frac15$&$\frac15$\\
$2\rule{0pt}{12pt}$&$\frac12$&$\frac13$&$\frac13$&$5$&$\frac23$&$\frac14$&
$\frac14$&
$8$&$\frac23$&$\frac15$&$\frac15$&$11$&$\frac25$&$\frac25$&$\frac25$&
$14$&$\frac12$&$\frac25$&$\frac13$\\
$3\rule{0pt}{12pt}$&$\frac23$&$\frac13$&$\frac13$&$6$&$\frac12$&$\frac13$&
$\frac15$&
$9$&$\frac12$&$\frac25$&$\frac15$&$12$&$\frac23$&$\frac13$&$\frac15$&
$15$&$\frac35$&$\frac25$&$\frac13$\\
&\rule{0pt}{2pt}&&&&&&&&&&&&&&&&&&\\
\hline
\end{tabular}}
\end{center}
\end{table}
In Table~\ref{tab:1} integers $p$ and $n$ are such that $2p\leq n$.
Further in the article instead of the scalar form of the hypergeometric
equation~(\ref{eq:euler}) we refer to its matrix form
\begin{equation}
 \label{eq:hyper_matrix}
  \frac{d}{dz}\Psi=\left(\frac{A_0}z+\frac{A_1}{z-1}\right)\Psi,
\end{equation}
where $A_0$ and $A_1\in sl_2(\mathbb{C})$. We also assume that
$A_0+A_1=-\frac{\theta_\infty}2\sigma_3$, with $\theta_\infty\in\mathbb{C}$
and $\sigma_3={\rm diag}\{1,-1\}$. Due to these conditions and the freedom in
normalization, $\Psi->\exp\{c\sigma_3\}\Psi$, $c\in\mathbb{C}$, the matrices 
$A_0$ and $A_1$, can be parametrized by the corresponding formal monodromies:
$\theta_0$, $\theta_1$, and $\theta_\infty$, where $\pm\theta_k/2$ are the
eigenvalues of the matrices $A_k$. Here we assume that 
{\bf$\theta_\infty\neq0$ and $A_0$ is not a diagonal matrix}. Under these
assumptions a fundamental solution $\Psi$ can be presented in terms of 
independent solutions of Equation~\ref{eq:euler}, for the parameters:
\begin{equation}
 \label{eq:theta-triples}
\lambda=\theta_\infty-1,\quad\mu=\theta_0,\quad\nu=\theta_1.
\end{equation}
This parametrization is not used here and therefore omitted; it can be found, 
say, in \cite{AK1}. In the following instead of the triples 
$(\lambda,\mu,\nu)$ we always
use $\theta$-triples, $(\theta_0,\theta_1,\theta_\infty)$, note $-1$
in Equations~(\ref{eq:theta-triples}). On the set of the $\theta$-triples
we assume the same equivalence relation as for $(\lambda,\mu,\nu)$ triples
and make no difference between the triples belonging to the same equivalence 
class.

The degenerate case of Equation~(\ref{eq:euler}) in the matrix framework is
described in the following way. Let the matrix $A_0$, and hence $A_1$, be 
triangular, but not diagonal. The lower triangular case corresponds to $a=0$ 
and the upper to $b=0$ in the parametrization considered in  \cite{AK1}. 
The triangular structure implies the following relation for the
$\theta$-triple: $\theta_0+\theta_1+\theta_\infty=0$. The general degenerate
case is obtained from the one of the triangular cases mentioned above by an
application of a finite number of Schlesinger transformations. 
These transformations change the $\theta$-triple such that they satisfy
the equation, $\theta_0+\theta_1+\theta_\infty=2k$ with some $k\in\mathbb{Z}$.
If $k\neq0$ the corresponding Equation~(\ref{eq:hyper_matrix}) does not have
a triangular structure, however its monodromy group remains isomorphic to the 
group of the triangular equation and has in particular the same triangular 
representation in a proper normalization. In the degenerate case the general
solution can be also algebraic in infinite number of cases they can be
deduced from the certain cases when the incomplete Beta function is
algebraic.   
    
Under our conditions Equation~(\ref{eq:hyper_matrix}) has three 
singular points at $0$, $1$, and $\infty$.
\begin{theorem}
 \label{Th:1}
All Equations~{\rm(\ref{eq:hyper_matrix})} corresponding to the parameters 
from the Schwarz List {\rm(}Table~{\rm\ref{tab:1})} can be obtained as 
inverse $RS$-transformations or compositions of $RS$-transformations and 
inverse $RS$-transformations of the $2\times2$ matrix Fuchsian ODE with two 
singular points,
\begin{equation}
 \label{eq:2Fuchs}
\frac{d}{dz}\Phi=\frac{A}z\Phi,\qquad{\rm where}\quad A\in sl_2(\mathbb C),
\end{equation}
has rational eigenvalues. Moreover, $R$-parts of the $RS$-transformations
are the Belyi functions.
\end{theorem}
{\it Proof of Theorem}~\ref{Th:1}.
We divide the 15 cases of Table~\ref{tab:1} into two sets; the first one 
contains all cases with $\lambda=1/2$ except the case $9$ and the second all 
the other cases. 
In Proposition~\ref{Prop:1/2} we show how to construct $RS$-transformations 
from Equation~(\ref{eq:hyper_matrix}) to Equation~(\ref{eq:2Fuchs}). 
This proves that every Equation~(\ref{eq:hyper_matrix}) corresponding to the 
first set is $RS$-inverse to Equation~(\ref{eq:2Fuchs}). 
In Proposition~\ref{Prop:non1/2} we prove that there exist 
$RS$-transformations from Equations~(\ref{eq:hyper_matrix}) corresponding
to the first set into Equations~(\ref{eq:hyper_matrix}) for the second one.
\begin{remark}{\rm
The fundamental solution of Equation~(\ref{eq:2Fuchs}) is 
$\Phi=Gz^{r\sigma_3}C$, where $G$ and $C\in SL(2,\mathbb{C})$,
and $r$ is a rational number. All $RS$-transformations that appear in
the following Propositions~(\ref{Prop:1/2})--(\ref{Prop:2-3}) can be
constructed explicitly, therefore, in fact, our proof provides an explicit 
construction of all general algebraic solutions of Equation~(\ref{eq:euler}).
}\end{remark}     
\begin{proposition}
 \label{Prop:1/2}
If a $\theta$-triple, defining Equation~{\rm(\ref{eq:hyper_matrix})}, 
coincides with one of $\theta$-triples corresponding to the raws of 
Table~{\rm\ref{tab:1}} with $\lambda=1/2$ except the raw $9$, i.e., 
the raws $1$, $2$, $4$, $6$, and $14$, then  there exists an 
$RS$-transformation, whose $R$-part is the Belyi function, which maps 
Equation~{\rm(\ref{eq:hyper_matrix})} to Fuchsian 
Equation~{\rm(\ref{eq:2Fuchs})}. 
\end{proposition}
{\it Proof.} Below we consider each of the 5 cases of 
Equation~{\rm(\ref{eq:hyper_matrix})} in the corresponding item. 
\begin{enumerate}
\item{Case 1.}
Actually, the general solution of Equation~(\ref{eq:euler})
for the triple $(1/2,\theta_1,1/2)$ with an arbitrary $\theta_1\in\mathbb{C}$ 
is an elementary function. This function is algebraic for rational values of 
$\theta_1$. These facts can be observed by application of transformation 
$RS_2^2\left(\!\!
\begin{tabular}{c|c|c}
$1/2$&$\theta_1$&$1/2$\\
$2$&$1+1$&$2$
\end{tabular}\!\!
\right)$ with the $R$-part $z=z_1^2$,
which maps Equation~{\rm(\ref{eq:hyper_matrix})} to
Equation~(\ref{eq:2Fuchs}).
\item{Case 2.} Here we use the transformation
$RS_2^2\left(\!\!
\begin{tabular}{c|c|c}
$1/3$&$1/2$&$1/3$\\
$3+1$&$2+2$&$3+1$
\end{tabular}\!\!
\right)$, with the Belyi function corresponding to the
truncated tetrahedron \cite{MZ}, $z=-\frac{64(z_1+1)^3}{(z_1(z_1-8))^3}$. 
\item{Case 4.} The required transformation is
$RS_2^2\left(\!\!
\begin{tabular}{c|c|c}
$1/4$&$1/2$&$1/3$\\
$4+1+1$&$2+2+2$&$3+3$
\end{tabular}\!\!
\right)$, with the Belyi function corresponding to the truncated cube
\cite{MZ}, $z=-\frac{108(z_1+1)^4z_1}{(z_1^2-14z_1+1)^3}$.
\item{Cases 6 and 14.} Transformations
$RS_2^2\left(\!\!\!\!
\begin{tabular}{c|c|c}
$1/3$&$1/2$&$1/5$\\
$\underbrace{3+\ldots+3}_4$&$\underbrace{2+\ldots+2}_6$&$5+5+1+1$
\end{tabular}\!\!\!\!
\right)$ and
$RS_2^2\left(\!\!\!\!
\begin{tabular}{c|c|c}
$1/3$&$1/2$&$2/5$\\
$\underbrace{3+\ldots+3}_4$&$\underbrace{2+\ldots+2}_{6}$&$5+5+1+1$
\end{tabular}\!\!\!\!
\right)$,
with the Belyi function of the truncated icosahedron \cite{MZ},
$z=\frac{(z_1^4+228z_1^3+494z_1^2-228z_1+1)^3}
{1728z_1(z_1^2-11z_1-1)^5}$.
\end{enumerate}
\begin{proposition}
 \label{Prop:non1/2}
Fundamental solutions of Equation~{\rm(\ref{eq:hyper_matrix})} corresponding
to $\theta$-triples for the raws: $3$, $5$, $7$, $8$, $9$, $10$, $11$, $12$, 
$13$, and $15$ of Table~{\rm\ref{tab:1}} can be constructed as 
$RS$-trans\-formations of the fundamental solutions of 
Equation~{\rm(\ref{eq:hyper_matrix})} corresponding to $\theta$-triples of 
the remaining raws of Table~{\rm\ref{tab:1}}, i.e., all the raws with 
$\lambda=1/2$ different from the $9$-th one. Moreover, $R$ parts of these 
transformations are the Belyi functions.   
\end{proposition} 
{\it Proof.}
At the end of item {\bf6.2} of \cite{AK1} it is shown that there are
quadratic, cubic, and sextic $RS$-transformations that map
Equation~(\ref{eq:hyper_matrix}) corresponding to the raw $6$ of 
Table~\ref{tab:1}
into the ones for the raws: $7$, $8$, $9$, $11$, $12$, and $13$.
At the end of item {\bf6.3} of the same work it is explained that using
quadratic, cubic, and sextic $RS$-transformations one can also map the 
``case'' $14$ of Table~\ref{tab:1} to the cases: $7$, $9$, $11$, $12$,
$13$, and $15$. To complete the proof we have to show how to obtain the cases
$3$, $5$, and $10$.

To get the case $3$ consider the quartic transformation,
$RS_3^2\left(\!\!
\begin{tabular}{c|c|c}
$\theta_0$&$1/2$&$1/4$\\
$2+1+1$&$2+2$&$4$
\end{tabular}\!\!
\right)$, 
where $\theta_0$ is arbitrary. An explicit form of the $R$ part
of this transformation, which is the Belyi function, can be found in item 
{\bf4.2.1.A} of \cite{AK2}. This $RS$ transformation can be also obtained 
as a composition of two quadratic transformations (cf.\cite{AK1}). 
The $\theta$-triple of the resulting Equation~(\ref{eq:hyper_matrix}) is 
$(\theta_0,\theta_0,2\theta_0-1)$. Thus choosing $\theta_0=1/3$ we get the 
case $3$ from the fourth one.

The Belyi function that allows one to build the $RS$-transformation from
the case $6$ of Table~\ref{tab:1} to the case $10$ is of the type   
$RS_3^2\left(\!\!
\begin{tabular}{c|c|c}
$1/3$&$1/2$&$1/5$\\
$3+3+2$&$\underbrace{2+\ldots+2}_4$&$5+2+1$
\end{tabular}\!\!
\right)$, 
its corresponding Belyi function reads
\begin{eqnarray}
 \label{eq:tr6-10}
&z=-\frac{(1000z_1^2-1728z_1+729)^3}{64(z_1-1)z_1^2(25z_1-27)^5},\quad
z-1=-\frac{(25000z_1^4-80000z_1^3+105300z_1^2-69984z_1+19683)^2}
{64(z_1-1)z_1^2(25z_1-27)^5}.&
\end{eqnarray}
It the dual function for the first dessin on Figure~\ref{fig:def521/332}.

The transformation  
$RS_3^2\left(\!\!
\begin{tabular}{c|c|c}
$1/3$&$1/2$&$1/4$\\
$4+3+3$&$\underbrace{2+\ldots+2}_5$&$4+4+1+1$
\end{tabular}\!\!
\right)$ maps the case $4$ to the case $5$.
Its $R$ part is the following Belyi function,
\begin{eqnarray}
 \label{eq:4-5}
\!\!\!\!\!\!\!
&z=\frac{(320z_1^2-320z_1-1)^3}{4z_1(z_1-1)(128z_1^2-128z_1+5)^4},\quad
z-1=-\frac{(2z_1-1)^2(16384z_1^4-32768z_1^3+15616z_1^2+768z_1-1)^2}
{4z_1(z_1-1)(128z_1^2-128z_1+5)^4}.&
\end{eqnarray}
\begin{remark}
 \label{rem:counterexample}{\rm
The Belyi function~(\ref{eq:tr6-10}) is easy to find with the help of
Maple 8. Although at first glance Function~(\ref{eq:4-5}) is only a 
little more complicated than Function~(\ref{eq:tr6-10}), I was not able 
to get it by analysing the corresponding system of algebraic equations 
(see Remark~\ref{Rem:m-parameter}) like it was done to find  
Function~(\ref{eq:tr6-10}); note that the factor $(2z_1-1)$ in the second 
Equation~(\ref{eq:4-5}) is not a priori evident, therefore the ansatz 
for its numerator is a square of the general polynomial of the fifth order 
with indeterminate coefficients. Moreover, there is a simpler ansatz for 
the function of degree $8$ with a proper number of the parameters that 
seems to produce a simpler Belyi function. It is Grothendieck's theory of 
``Dessin's d'Enfants'' that helped to find a correct degree and symmetry 
of the Belyi function (see explanation on Figure~\ref{fig:4-5}).}
\end{remark}
\begin{figure}[ht]
\begin{picture}(30,50)
\put(30,25){\circle{30}}
\put(70,25){\circle{30}}
\put(30,25){\line(1,0){40}}
\put(30,25){\circle*{3}}
\put(70,25){\circle*{3}}
\put(45,25){\circle*{3}}
\put(55,25){\circle*{3}}
\put(100,25){$\Longrightarrow\quad
z=\frac{500}{729}\frac{(z_1^2-6/5)^3}{(z_1^2-1)^4(z_1^2-32/27)},
\quad
z-1=-\frac{z_1^2(z_1^4-70/27z_1^2+5/3)^2}
{(z_1^2-1)^4(z_1^2-32/27)}.
$}
\end{picture}
\caption{The dessin for the Belyi function of the type
$R(4+3+3|2+2+2+2+2|4+4+1+1)$.}
  \label{fig:4-5}
\end{figure}
\begin{remark}{\rm
Of course, there are some other transformations that map  
Equations~(\ref{eq:hyper_matrix}) corresponding to different cases of 
Table~\ref{tab:1} to each other including some autotransformations.
In Proposition~\ref{Prop:2-3} we give an example of the transformation
of the case 2 to the case 3 of Table~\ref{tab:1}. Due to the appearance
of an arbitrary parameter $s$, this transformation allows one to get a 
fundamental solution of Equation~(\ref{eq:hyper_matrix}) for the case 3 
in terms of the corresponding Schlesinger transformation without usage 
of the explicit form of the initial fundamental solution of 
Equation~(\ref{eq:hyper_matrix}) for the case 2.
}\end{remark}
\begin{proposition}
 \label{Prop:2-3}
The transformation
$RS_3^2\!\left(\!\!\!\!
\begin{tabular}{c|c|c}
$1/3$&$1/2$&$1/3$\\
$3+1+1+1\!$&$\!2+2+2\!$&$\!3+3$
\end{tabular}\!\!\!\!
\right)$ 
maps Equation~{\rm(\ref{eq:hyper_matrix})} corresponding to the third
raw of Table~{\rm\ref{tab:1}} to Equation~{\rm(\ref{eq:hyper_matrix})} 
for the second raw. 
\end{proposition}
{\it Proof.}
For the proof it is enough to present the rational function
of the type $R(3+1+1+1|2+2+2|3+3)$, which is given by
Equation~(\ref{eq:3111itd}). 
\section{One Irreducible Octic Transformation}
 \label{sec:non-Schwarz}
It is indicated in \cite{AK1} that Belyi function of the following type
$R(7+1|2+2+2+2|3+3+1+1)$, i.e.,  
\begin{equation}
 \label{eq:tr8}
z=\frac{\rho z_1(z_1-a)^7}
{(z_1-1)(z_1-c_1)^3(z_1-c_2)^3},\;\;
z-1=\frac{\rho(z_1-b_1)^2(z_1-b_2)^2(z_1-b_3)^2(z_1-b_4)^2}
{(z_1-1)(z_1-c_1)^3(z_1-c_2)^3}\!,
\end{equation}
defines three irreducible octic transformations of the hypergeometric 
function, which in terms of $\theta$-triples read,
\begin{equation}
 \label{eq:rs-octic}
\left(\frac12,\frac13,\frac17\right)\to\left(\frac13,\frac23,\frac17\right),
\quad
\left(\frac12,\frac13,\frac27\right)\to\left(\frac13,\frac13,\frac27\right),
\quad
\left(\frac12,\frac13,\frac37\right)\to\left(\frac13,\frac23,\frac37\right).
\end{equation}
In \cite{AK1} Function~(\ref{eq:tr8}) was not identified as the Belyi 
function, therefore its existence was not strictly speaking established. 
However, we were able there to calculate with a very high accuracy its 
coefficients numerically. Now, the existence of this transformation follows 
from the dessin presented on the picture below.\vspace{8pt}\\  
\parbox{100pt}
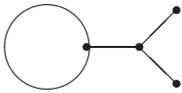
{
\begin{picture}(30,35)
 \label{fig:8}
\put(30,15){\circle{30}}
\put(45,15){\line(1,0){20}}
\put(65,15){\line(1,1){14}}
\put(65,15){\line(1,-1){14}}
\put(79,29){\circle*{3}}
\put(65,15){\circle*{3}}
\put(45,15){\circle*{3}}
\put(79,01){\circle*{3}}
\end{picture}}
\parbox{320pt}{
Moreover, Maple 8 and a substantially better computer than that used in 
calculations for \cite{AK1} allow to find them now explicitly.
To present the result we rewrite function~(\ref{eq:tr8}) in the 
following form:}
\begin{equation}
 \label{eq:tr8mod}
z=\frac{\rho z_1(z_1-a)^7}
{(z_1-1)(z_1^2+\hat c_1z_1+\hat c_0)^3},\;\;
z-1=\frac{\rho(z_1^4+\hat b_3z_1^3+\hat b_2z_1^2+\hat b_1z_1+\hat b_0)^2}
{(z_1-1)(z_1^2+\hat c_1z_1+\hat c_0)^3}.
\end{equation}
where
\begin{eqnarray}
 \label{eq:8rho-a-c}
&\rho=\frac{1-i3\sqrt{3}}{112},\quad a=\frac{27-i39\sqrt{3}}{98},\quad
\hat c_0=\frac{-5697+i2349\sqrt{3}}{268912},\quad
\hat c_1=-\frac{513+i435\sqrt{3}}{784},\phantom{aa}&\\
\label{eq:8b}
&\!\hat b_0=\frac{60507+i142803\sqrt{3}}{13176688},\;\;
\hat b_1=\frac{249399-i38313\sqrt{3}}{134456},\;\;
\hat b_2=\frac{-4293+i28251\sqrt{3}}{5488},\;\;
\hat b_3=-\frac{83+i129\sqrt{3}}{28}.&\phantom{aa}
\end{eqnarray}
It is, of course, straightforward to find explicit formulae for the roots:
$b1$, $b_2$, $b_3$, $b_4$ and $c_1$, $c_2$. We omit them as the ones for
$b_k$, $k=1,2,3,4$ are very cumbrous. Moreover, corresponding 
$RS$-transformations are actually symmetric functions of these roots, 
i.e., they can be expressed in terms of the 
coefficients~(\ref{eq:8rho-a-c}) and (\ref{eq:8b}).

Each of the octic transformations~(\ref{eq:rs-octic}) together with the
quadratic, cubic, and their inverse ones (cf. \cite{K1}) generates a cluster 
of the hypergeometric functions which have the same type of transcendency.
We have presented these clusters in terms of the corresponding 
$\theta$-triples in Table~\ref{tab:2}.
\par
\begin{table}[ht]
 \caption{Three Octic Clusters}
  \label{tab:2}
\begin{center}
{\renewcommand{\arraystretch}{0}%
\begin{tabular}{|c|ccc||c|ccc||c|ccc||c|ccc|}
\hline
\multicolumn{16}{|c|}{The First Octic Cluster\phantom{$\frac{A}{B}$}}\\
\hline
\strut$N$&$\theta_0$&$\theta_1$&$\theta_\infty$&$N$&$\theta_0$&$\theta_1$&
$\theta_\infty$&$N$&$\theta_0$&$\theta_1$&$\theta_\infty$&$N$&$\theta_0$&
$\theta_1$&$\theta_\infty$\\
\hline
$1\rule{0pt}{12pt}$&$\frac12$&$\frac13$&$\frac 17$&$4$&$\frac12$&$\frac17$&
$\frac1{14}$&$7$&$\frac13$&$\frac17$&$\frac17$&$10$&$\frac67$&$\frac17$&$
\frac17$\\
$2\rule{0pt}{12pt}$&$\frac12$&$\frac13$&$\frac1{14}$&$5$&$\frac13$&$\frac13$&
$\frac57$&$8$&$\frac13$&$\frac1{14}$&$\frac1{14}$&$11$&$\frac57$&$\frac27$&
$\frac27$\\
$3\rule{0pt}{12pt}$&$\frac12$&$\frac27$&$\frac17$&$6$&$\frac13$&$\frac13$&
$\frac67$&$9$&$\frac37$&$\frac17$&$\frac17$&$12$&$\frac57$&$\frac1{14}$&
$\frac1{14}$\\
&\rule{0pt}{2pt}&&&&&&&&&&&&&&\\
\hline
\multicolumn{16}{|c|}{The Second Octic Cluster\phantom{$\frac{A}{B}$}}\\
\hline
\strut$N$&$\theta_0$&$\theta_1$&$\theta_\infty$&$N$&$\theta_0$&$\theta_1$&
$\theta_\infty$&$N$&$\theta_0$&$\theta_1$&$\theta_\infty$&$N$&$\theta_0$&
$\theta_1$&$\theta_\infty$\\
\hline
$1\rule{0pt}{12pt}$&$\frac12$&$\frac13$&$\frac27$&$4$&$\frac12$&$\frac27$&
$\frac5{14}$&$7$&$\frac13$&$\frac27$&$\frac27$&$10$&$\frac37$&$\frac5{14}$&
$\frac5{14}$\\
$2\rule{0pt}{12pt}$&$\frac12$&$\frac13$&$\frac5{14}$&$5$&$\frac13$&$\frac13$&
$\frac37$&$8$&$\frac13$&$\frac5{14}$&$\frac5{14}$&$11$&$\frac27$&$\frac27$&
$\frac27$\\
$3\rule{0pt}{12pt}$&$\frac12$&$\frac27$&$\frac37$&$6$&$\frac13$&$\frac13$&
$\frac27$&$9$&$\frac37$&$\frac37$&$\frac37$&$12$&$\frac27$&$\frac27$&
$\frac17$\\
&\rule{0pt}{2pt}&&&&&&&&&&&&&&\\
\hline
\multicolumn{16}{|c|}{The Third Octic Cluster\phantom{$\frac{A}{B}$}}\\
\hline
\strut$N$&$\theta_0$&$\theta_1$&$\theta_\infty$&$N$&$\theta_0$&$\theta_1$&
$\theta_\infty$&$N$&$\theta_0$&$\theta_1$&$\theta_\infty$&$N$&$\theta_0$&
$\theta_1$&$\theta_\infty$\\
\hline
$1\rule{0pt}{12pt}$&$\frac12$&$\frac13$&$\frac37$&$4$&$\frac12$&$\frac37$&
$\frac17$&$7$&$\frac13$&$\frac37$&$\frac37$&$10$&$\frac47$&$\frac37$&$\frac27$
\\
$2\rule{0pt}{12pt}$&$\frac12$&$\frac13$&$\frac3{14}$&$5$&$\frac13$&$\frac13$&
$\frac47$&$8$&$\frac13$&$\frac3{14}$&$\frac3{14}$&$11$&$\frac17$&$\frac17$&
$\frac17$\\
$3\rule{0pt}{12pt}$&$\frac12$&$\frac37$&$\frac3{14}$&$6$&$\frac13$&$\frac13$&
$\frac17$&$9$&$\frac47$&$\frac37$&$\frac37$&$12$&$\frac17$&$\frac3{14}$&
$\frac3{14}$\\
&\rule{0pt}{2pt}&&&&&&&&&&&&&&\\
\hline
\end{tabular}}
\end{center}
\end{table} 
Note that the first cluster contains the hypergeometric functions 
corresponding to the triples 
$\big(\frac57,\frac57,\frac57\big)$ $\&$ 
$\big(\frac67,\frac67,\frac67\big)$, 
the second --
$\big(\frac27,\frac27,\frac27\big)$ $\&$ 
$\big(\frac37,\frac37,\frac37\big)$, 
and the third --
$\big(\frac17,\frac17,\frac17\big)$ $\&$ 
$\big(\frac47,\frac47,\frac47\big)$. 
\appendix
 \section{Appendix. On Quadratic Transformations for the Sixth Painlev\'e 
Equation}
 \label{app:A}
Existence of the quadratic transformations for the sixth Painlev\'e
equation were discovered in \cite{K3} via an artificial transformation
found for the similarity reductions of the so-called three-wave resonant
system. In the subsequent work \cite{K4} one quadratic transformation
was derived explicitly via the method of $RS$ transformations.
The latter transformation acts on the $\theta$-tuples as,
\begin{equation}
 \label{eq:quadratic-kit}
\left(\frac12,\hat\theta_1^0,\hat\theta_t^0,\pm\frac12\right)\quad
\longmapsto\quad
\left(\hat\theta_t^0,\hat\theta_1^0,\hat\theta_t^0,\hat\theta_1^0\right).
\end{equation}
For the solution $y(t)$ corresponding to the right $\hat\theta$-tuple 
and depending on $t=4\sqrt{t_0}/(1+\sqrt{t_0})^2$, where $t_0$ is
the independent variable for the original solution $y_0(t_0)$
corresponding to the left $\hat\theta$-tuple, it was obtained an explicit 
though complicated formula in terms of $y_0(t_0)$.  
Recently K. Okamoto and his collaborators found that this formula can be 
actually substantially simplified. Some time ago Manin \cite{M}
rediscovered the so-called elliptic form of the sixth Painlev\'e equation
given first by Painlev\'e. He applied then the Landen transformation for 
the elliptic functions and found another transformation in terms of the 
``elliptic variables'' for the sixth Painlev\'e equation. It was later 
mentioned by Okamoto that it is also a quadratic transformation however, it 
is not equivalent to the one given above. Actually, Manin's transformation 
cannot be obtained via the method of $RS$-transformation if we apply it to 
the associated linear Fuchsian $2\times2$ matrix ODE in the Jimbo-Miwa 
parametrization \cite{JM}; in this parametrization we get only 
quadratic transformation~(\ref{eq:quadratic-kit}) and its equivalent forms. 
The Jimbo-Miwa parametrization, which is proved to be a very helpful one for 
studies of almost all questions related with the theory of the sixth 
Painlev\'e equation, has one drawback: the complete group of symmetries for 
this equation \cite{O} cannot be realized as the group action on the 
solutions of the Fuchsian ODE. Therefore if we want to build the whole
theory for the sixth Painlev\'e equation based on the isomonodromy problem 
for the ($2\times2$) matrix linear Fuchsian ODE, then for studies of some 
questions, especially the ones related with the symmetries and 
transformations, we have to consider alternatively another parametrization 
of this Fuchsian ODE by employing one special transformation found by 
Okamoto~\cite{O}. We call the last transformation the 
{\it Okamoto transformation} and related parametrization the 
{\it Okamoto parametrization} (see Appendix \cite{KK} 
Equation~(A.5)\footnote{There is a misprint in Equation~(A.6) of \cite{KK}; 
instead of $y_0^2$ in the numerator there should be $y_0$.}). Recently, in 
the work \cite{NTY} a representation of the complete group of Okamoto's 
transformations as the symmetries for solutions of a linear ODE in matrix 
dimension greater than 2 has been obtained.

If we apply the method of $RS$-transformation to the $2\times2$ Fuchsian
ODE with four singular points in the Okamoto parametrization, then we can 
obtain the Manin transformation for the canonical form of the sixth
Painlev\'e equation. Here I will not give an explicit form of the Okamoto 
parametrization and the ``$RS$-derivation'' of the Manin transformation. 
Instead I present the final answer in the soul close to the presentation in 
the main part of this paper.
  
Now we describe an algorithm how to construct quadratic transformations of 
the Manin type. Consider the quadratic Belyi functions, 
$z(z_1)$ mapping $\mathbb{CP}^1$ into itself:
\begin{equation}
 \label{eq:quadratic-belyi}
z=z_1^2,\quad 
z=-4z_1(z_1-1),\quad
z=\frac{z_1^2}{4(z_1-1)}.
\end{equation}
\begin{remark}{\rm
These functions, of course, have only two critical points on 
the Riemann sphere. However, Proposition~\ref{Prop:necessary_condition}
is applicable for them too: if we formally put the multiplicity of the
absent third critical point equal zero, then it gives $m=0$, as it should 
be for the Belyi functions.}
\end{remark} 
Here we study transformations of the type: $RS_4^2(4)$, therefore we 
incorporate in our notation one more point 
$t\in \mathbb{CP}^1\setminus\{0,1,\infty\}$. Thus the types of these 
functions read, respectively, as follows: $R(2|1+1|1+1|2)$, 
$R(|1+1|2|1+1|2)$, and $R(2|2|1+1|1+1)$, where the third box is for $t$
and the last one for the point at $\infty$.  
\begin{remark}{\rm
In this case it is also convenient to treat all four points $0$, $1$, $t$, 
$\infty$ on ``equal footing'' rather than normalize the critical values of 
the Belyi functions  necessarily at $0$, $1$, and $\infty$. Finally the 
latter ambiguity is absorbed by the corresponding transformations for the 
sixth Painlev\'e equation. However, instead of making these transformations 
afterwards it is easier to prepare the desired quadratic transformation from 
the very beginning as it does not require any special efforts. For example, 
we consider also the function $z=1+(1-t)(z_1-1)^2/(4z_1)$, which is of the 
type $R(1+1|2|2|1+1)$. Clearly, we have $6$ different quadratic Belyi 
functions.
}\end{remark}
For a given type of the Belyi function consider the $\hat\theta$-tuple whose
members are denoted as $\hat\theta_k^0$, $k=0,1,t_0,\infty$. The parameters 
$\hat\theta_k^0$ corresponding to the boxes of the type symbol with $2$ 
vanish if $k=0,1,t_0$, or $\hat\theta_k^0=1$ for $k=\infty$, the other
two members of the $\hat\theta$-tuple are arbitrary. Now, denote $y_0(t_0)$
any solution of the sixth Painlev\'e equation~(\ref{eq:P6}) with $t=t_0$ 
and the $\hat\theta$-tuple defined above. The set of the preimages 
$\{z^{-1}(0),z^{-1}(1),z^{-1}(t_0),z^{-1}(\infty)\}$ contains four
non-apparent points, namely, these are preimages of the points with $1+1$ 
boxes in the type of $z(z_1)$. Denote $\tilde z(z_1)$ any of 24 
fractional linear transformations which map the set of non-apparent 
preimages of $z(z_1)$ into sets of the form $\{0,1,t,\infty\}$, with some $t$.
\begin{proposition}
 \label{Prop:quadratic}
Let $\tau\in\{z^{-1}(0),\,z^{-1}(1),\,z^{-1}(t_0),\,z^{-1}(\infty)\}$ be 
non-apparent and such that $\tilde z(\tau)\not\in\{0,1,\infty\}$. Define
$$
t=\tilde z(\tau),\qquad 
y(t)=\tilde z(z^{-1}(y_0(t_0))),\qquad i_k=z(\tilde z^{-1}(k)),\qquad
k=0,1,t,\infty. 
$$
The function $y(t)$ solves the sixth Painlev\'e equation for 
the $\hat\theta$-tuple with the members 
$
\hat\theta_k=\frac12\hat\theta^0_{i_k}
$
if $k\neq\infty$ and $i_k\neq\infty$,
$\hat\theta_k=\frac12(\hat\theta^0_\infty-1)$ if $k\neq\infty$ and 
$i_k=\infty$, and $\hat\theta_\infty-1=\frac12(\hat\theta^0_\infty-1)$.
\end{proposition}
\begin{remark}{\rm
All construction is completely invertible so that for every quadratic
transformation we can define the inverse transformation which is also
called quadratic transformation. All in all this construction gives
$6\cdot24=144$ quadratic transformations without counting their inverses. 
However, of course, if we consider them modulo fractional linear 
transformations interchanging $0,1,t,\infty$ for both the initial equation 
and for final one, then we have actually only two seed not equivalent 
transformations, since there is a difference between two cases, namely, 
depending on whether $\infty$ is set as a critical value of the Belyi 
function or not. If we further factorize them modulo 
B\"acklund transformations, then we left with only one seed transformation. 
This seed transformation can be derived from the quadratic transformation
found by the author (\ref{eq:quadratic-kit}) via application of the Okamoto 
transformation. Therefore, all in all there is only one seed quadratic 
transformation for the sixth Painlev\'e equation. However, in those cases 
when the explicit formula is needed it is convenient to use directly 
Proposition~\ref{Prop:quadratic} as it allows, in many situations, to avoid 
compositions of a cumbersome transformations.
}\end{remark}
Below we consider examples of quadratic transformations constructed by
means of Proposition~\ref{Prop:quadratic} for the Belyi 
functions~(\ref{eq:quadratic-belyi}). Example 3 is the algebraic form of
the transformation obtained by Manin.
\begin{eqnarray*}
1.\qquad\hspace{1.5cm} z(z_1)=z_1^2,&&
\tilde z(z_1)=\frac2{1+\sqrt{t_0}}\frac{z_1+\sqrt{t_0}}{z_1+1},
\quad\tau=\sqrt{t_0},\hspace{2.8cm}\\
t=\frac{4\sqrt{t_0}}{(1+\sqrt{t_0})^2},&&
y(t)=\frac2{1+\sqrt{t_0}}
\frac{\sqrt{y_0(t_0)}+\sqrt{t_0}}{\sqrt{y_0(t_0)}+1},\\
\left(0,\hat\theta_1^0,\hat\theta_{t_0}^0,1\right)&\longmapsto&
\left(\frac12\hat\theta_{t_0}^0,\frac12\hat\theta_1^0,
\frac12\hat\theta_{t_0}^0,1+\frac12\hat\theta_1^0\right).\\
\end{eqnarray*}
\begin{eqnarray*}
2.\qquad z(z_1)=-4z_1(z_1-1),&&
\tilde z(z_1)=\frac{\left(1-\sqrt{1-t_0}\right)z_1}{2z_1-1-\sqrt{1-t_0}},
\quad\tau=\frac12\left(1-\sqrt{1-t_0}\right),\\
t=-\frac{\left(1-\sqrt{1-t_0}\right)^2}{4\sqrt{1-t_0}},&&
y(t)=-\frac{\left(1-\sqrt{1-t_0}\right)\left(1-\sqrt{1-y_0(t_0)}\right)}
{2\left(\sqrt{1-t_0}+\sqrt{1-y_0(t_0)}\right)},\\
\left(\hat\theta_0^0,0,\hat\theta_{t_0}^0,1\right)&\longmapsto&
\left(\frac12\hat\theta_0^0,\frac12\hat\theta_0^0,
\frac12\hat\theta_{t_0}^0,1+\frac12\hat\theta_{t_0}^0\right).\\
\end{eqnarray*}
\begin{eqnarray*}
3.\qquad z(z_1)=\frac{z_1^2}{4(z_1-1)},\!\!&\!\!&\!\!\!
\tilde z(z_1)=\frac{\left(\!z_1-2t_0+2\sqrt{t_0^2-t_0}\right)}
{\left(\!1-2t_0+2\sqrt{t_0^2-t_0}\right)},
\;\tau=2\left(\!t_0+\sqrt{t_0^2-t_0}\right)\!,\\
t=\frac{4\sqrt{t_0^2-t_0}}
{\left(1-2t_0+2\sqrt{t_0^2-t_0}\right)},\!\!\!&\!\!&
y(t)=\frac{2\left(y_0(t_0)+\sqrt{y_0(t_0)^2-y_0(t_0)}-t_0+
\sqrt{t_0^2-t_0}\right)}{\left(1-2t_0+2\sqrt{t_0^2-t_0}\right)},\\
\left(0,0,\hat\theta_{t_0}^0,\hat\theta_\infty^0\right)\!\!&\!
\longmapsto\!&\!\!
\left(\frac12\hat\theta_{t_0}^0,\frac12\big(\hat\theta_\infty^0-1\big),
\frac12\hat\theta_{t_0}^0,1+\frac12\big(\hat\theta_\infty^0-1\big)\right).
\end{eqnarray*}
\begin{remark}{\rm
In all formulae 1--3 above the branches of the square roots with
$y_0(t_0)$ can be taken independently of the branches of square roots
with $t_0$; it does not change the mapping between the $\hat\theta$-tuples.
All transformations are invertible}
\end{remark}

\end{document}